\documentclass{article}
\usepackage{amssymb}
\usepackage{amsmath}

\input{tcilatex}
\begin{document}

\begin{center}
416ArXivFiniteDimReprDifferenceOpREV141205

\bigskip

{\Huge Finite-dimensional representations of difference operators, and the
identification of remarkable matrices}

\bigskip

{\Large Francesco Calogero}

Physics Department, University of Rome "La Sapienza", 00185 Rome, Italy

Istituto Nazionale di Fisica Nucleare, Sezione di Roma

\bigskip

\textit{Abstract}
\end{center}

Two square matrices of (arbitrary) order $N$ are introduced. They are
defined in terms of $N$ arbitrary numbers $z_{n}$, and of an arbitrary
additional parameter ($a$ respectively $q$), and provide finite-dimensional
representations of the two operators acting on a function $f(z)$ as follows: 
$[f(z+a)-f(z)]/a$ respectively $[f(qz)-f(z)]/[(q-1)z]$. These
representations are exact---in a sense explained in the paper---when the
function $f(z)$ is a polynomial in $z$ of degree less than $N$. This
formalism allows to transform difference equations valid in the space of
polynomials of degree less than $N$ into corresponding matrix-vector
equations. As an application of this technique several remarkable square
matrices of order $N$ are identified, which feature explicitly $N$ arbitrary
numbers $z_{n}$, or the $N$ zeros of polynomials belonging to the Askey and $%
q$-Askey schemes. Several of these findings have a Diophantine
character.\bigskip

\pagebreak

\section{Introduction}

This paper is focussed on $\left( N\times N\right) $-matrices which provide
finite-dimensional representations of difference operators yielding \textit{%
exact} results in the context of the functional space spanned by polynomials
of degree less than $N$; the precise meaning of this statement is clarified
below.

These findings extend to \textit{difference} operators the results reported
for the standard \textit{differential} operator in Section 2.4 (entitled
"Finite dimensional representations of differential operators, Lagrangian
interpolation, and all that") of \cite{C2001} (and see also papers referred
to there). Let us summarize here---for completeness, and also to introduce
some notation used throughout---the essence of those findings.

\textit{Notation 1.1}. Throughout this paper $N$ is an arbitrary positive
integer (unless otherwise explicitly indicated); $N$-vectors are denoted by
underlined (Latin or Greek) letters, so that, for instance, the $N$ vector $%
\underline{v}$ has the $N$ components $v_{n};$ likewise $\left( N\times
N\right) $-matrices are denoted by twice-underlined (Latin or Greek)
letters, so that, for instance, the $\left( N\times N\right) $-matrix $%
\underline{\underline{M}}$ features the $N^{2}$ components $M_{nm}.$ Here
and throughout the indices $n,$ $m,$ $\ell $ run over the integers from $1$
to $N$, unless otherwise indicated. Attention is generally restricted to
functions $f\left( z\right) $ which depend \textit{analytically} on their
argument $z,$ and in particular that are polynomials in their argument $z$.
The formulas written below are generally valid for arbitrary, complex values
of all the quantities denoted by (Latin or Greek) letters, up to obvious
limitations for cases when limits might have to be taken: for instance $%
g\left( a,z\right) =\left[ f\left( a+z\right) -f\left( z\right) \right] /a$
has a clear significance for every value of the quantity $a$ except for $a=0$%
, but it also clearly implies $g\left( 0,z\right) =df\left( z\right) /dz$.
Finally, we use throughout the notation $\mathbf{i}$ to denote the \textit{%
imaginary unit}, so that $\mathbf{i}^{2}=-1$. $\square $

To summarize the previous results \cite{C2001} let us assume that the
function $f\left( z\right) $ is a polynomial of degree less than $N$,%
\begin{equation}
f\left( z\right) =\sum_{m=1}^{N}\left[ c_{m}~z^{N-m}\right] ~,  \label{Polf}
\end{equation}%
and let us then express it as a linear combination---with coefficients $%
f_{n} $---of the $N$ interpolational polynomials $p_{N-1}^{\left( n\right)
}\left( z\right) $---all of them of degree $N-1$ in $z$---defined as follows
in terms of the $N$, \textit{arbitrarily assigned}, numbers $z_{n}$: 
\begin{equation}
p_{N-1}^{\left( n\right) }\left( z\right) \equiv p_{N-1}^{\left( n\right)
}\left( z;\underline{z}\right) =\prod\limits_{\ell =1,~\ell \neq
n}^{N}\left( \frac{z-z_{\ell }}{z_{n}-z_{\ell }}\right) ~,
\label{InterpolPol}
\end{equation}%
\begin{equation}
f\left( z\right) =\sum_{n=1}^{N}\left[ f_{n}~p_{N-1}^{\left( n\right)
}\left( z\right) \right] ~.  \label{ffn}
\end{equation}%
This definition, (\ref{InterpolPol}), of the interpolational polynomials $%
p_{N-1}^{\left( n\right) }\left( z\right) $ clearly implies the relation%
\begin{equation}
p_{N-1}^{\left( n\right) }\left( z_{m}\right) \equiv p_{N-1}^{\left(
n\right) }\left( z_{m};\underline{z}\right) =\delta _{nm}~,~~~n,m=1,...,N~;
\end{equation}%
hence the $N$ coefficients $f_{n}$ in the right-hand side of (\ref{ffn}) are
just the values of $f\left( z\right) $ at the $N$ points $z_{n}$:%
\begin{equation}
f_{n}=f\left( z_{n}\right) ~,~~~n=1,...,N~.
\end{equation}

The above formulas correspond to the standard formulation of Lagrangian
interpolation: the $N$ interpolational points $z_{n}$ can be arbitrarily
assigned, except for the restriction that they be all different among
themselves, see (\ref{InterpolPol}) (otherwise some limits would have to be
taken). They entail a one-to-one relationship among the $N$-vector $%
\underline{f}$ featuring the $N$ components $f_{n}$,%
\begin{equation}
\underline{f}=\left( f_{1},...,f_{N}\right) ~,
\end{equation}%
and the function $f\left( z\right) $ restricted to be a polynomial in $z$ of
degree less than $N$.

In Section 2.4 of \cite{C2001} certain relations are reported among \textit{%
differential} operators acting on such functions $f\left( z\right) ,$ see (%
\ref{Polf}), and appropriately defined $\left( N\times N\right) $-matrices
acting on the $N$-vector $\underline{f}$. The basic formula of this kind,
corresponding to the definition 
\begin{equation}
f^{\left\langle r\right\rangle }\left( z\right) =\left( \frac{d}{dz}\right)
^{r}~f\left( z\right) ~,~~~r=0,1,2,...~,  \label{rDerf}
\end{equation}%
reads (see, up to minor notational changes, eq. (2.4.1-9) of \cite{C2001}) 
\begin{equation}
\underline{f}^{\left\langle r\right\rangle }=\left( \underline{\underline{V}}%
~\underline{\underline{D}}~\underline{\underline{V}}^{-1}\right) ^{r}~%
\underline{f}=\underline{\underline{V}}~\left( \underline{\underline{D}}%
\right) ^{r}~\underline{\underline{V}}^{-1}~\underline{f}=\underline{%
\underline{V}}~\left( \underline{\underline{D}}\right) ^{r}~f\left( 
\underline{\underline{Z}}\right) ~\underline{w}~,~~~r=0,1,2,...~.
\label{VecDerfr}
\end{equation}%
Here the $N$ components $f_{n}^{\left\langle r\right\rangle }$ of the $N$%
-vector$\underline{f}^{\left\langle r\right\rangle }$ are of course the $N$
values $f^{\left\langle r\right\rangle }\left( z_{n}\right) $ that the $r$%
-th derivative $f^{\left\langle r\right\rangle }\left( z\right) $ of the
function $f\left( z\right) $ (see (\ref{rDerf})) takes at the $N$
interpolational points $z_{n}$, 
\begin{subequations}
\begin{equation}
f_{n}^{\left\langle r\right\rangle }\equiv \left( \underline{f}%
^{\left\langle r\right\rangle }\right) _{n}=f^{\left\langle r\right\rangle
}\left( z_{n}\right) ~,~~\ n=1,...,N~,~~~r=0,1,2,...
\end{equation}%
or, equivalently, 
\begin{equation}
\underline{f}^{\left\langle r\right\rangle }=f^{\left\langle r\right\rangle
}\left( \underline{\underline{Z}}\right) ~\underline{u}~,
\end{equation}%
with the $N$-vector $\underline{u}$ having all elements equal to \textit{%
unity}, 
\end{subequations}
\begin{equation}
\underline{u}=\left( 1,~1,...,~1\right) ~;~~~u_{n}=1~,~~~n=1,...,N~.
\label{Vecu}
\end{equation}%
The $\left( N\times N\right) $-matrices $\underline{\underline{Z}}\equiv 
\underline{\underline{Z}}\left( \underline{z}\right) $~, $\underline{%
\underline{D}}\equiv \underline{\underline{D}}\left( \underline{z}\right) $, 
$\underline{\underline{V}}\equiv \underline{\underline{V}}\left( \underline{z%
}\right) $ and the $N$-vector $\underline{w}\equiv \underline{w}\left( 
\underline{z}\right) $ are defined componentwise as follows, in terms of the 
$N$ arbitrary numbers $z_{n}$ which are the $N$ components of the $N$-vector 
$\underline{z}$:%
\begin{equation}
\underline{\underline{Z}}\left( \underline{z}\right) =\text{diag}\left[ z_{n}%
\right] ~;~~~Z_{nm}=\delta _{nm}~z_{n}~;  \label{MatZ}
\end{equation}%
\begin{subequations}
\label{MatD}
\begin{equation}
D_{nn}\equiv D_{nn}\left( \underline{z}\right) =\sum_{\ell =1,~\ell \neq
n}^{N}\left( \frac{1}{z_{n}-z_{\ell }}\right) ~,~~~n=1,...N~,  \label{MatDnn}
\end{equation}%
\begin{equation}
D_{nm}\equiv D_{nm}\left( \underline{z}\right) =\left( \frac{1}{z_{n}-z_{m}}%
\right) ~,~~~n\neq m,~~~n,m=1,...N~;  \label{MatDnm}
\end{equation}%
\end{subequations}
\begin{equation}
V_{nm}\equiv V_{nm}\left( \underline{z}\right) =\delta
_{nm}~\prod\limits_{\ell =1,~\ell \neq n}^{N}\left( z_{n}-z_{\ell }\right) ~;
\label{MatV}
\end{equation}%
\begin{equation}
\underline{w}=\underline{V}^{-1}~\underline{u}~,~~~w_{n}\equiv w_{n}\left( 
\underline{z}\right) =\prod\limits_{\ell =1,~\ell \neq n}^{N}\left( \frac{1}{%
z_{n}-z_{\ell }}\right) ~.  \label{Vecw}
\end{equation}

These formulas imply \cite{C2001} that, whenever a function $f\left(
z\right) $ which is a polynomial of degree less than $N$ in $z$, see (\ref%
{Polf}), satisfies the differential equation (with no \textit{a priori}
restriction on the positive integer $R$ and on the $R+2$ functions $%
d_{r}\left( z\right) $ and $g\left( z\right) $) 
\begin{subequations}
\begin{equation}
\sum_{r=0}^{R}\left[ d_{r}\left( z\right) ~\left( \frac{d}{dz}\right)
^{r}~f\left( z\right) \right] =g\left( z\right) ~,
\end{equation}%
the following $N$-vector equation is as well valid:%
\begin{equation}
\sum_{r=0}^{R}\left[ d_{r}\left( \underline{\underline{Z}}\right) ~\left( 
\underline{\underline{D}}\right) ^{r}\right] ~f\left( \underline{\underline{Z%
}}\right) ~\underline{w}=g\left( \underline{\underline{Z}}\right) ~%
\underline{w}~;
\end{equation}%
with the $\left( N\times N\right) $-matrices $\underline{\underline{Z}}$ and 
$\underline{\underline{D}}$ and the $N$-vector $\underline{w}$ defined as
above, see (\ref{MatZ}), (\ref{MatD}) and (\ref{Vecw}), in terms of $N$
numbers $z_{n}$, arbitrary except for the condition that they be all
different among themselves. With remarkable consequences \cite{C2001}.

The main findings of this paper extend to \textit{difference} operators
these results; they are reported in the next Section 2 and proven in Section
4. In Section 3, as examples of applications of these findings several $%
\left( N\times N\right) $-matrices displaying remarkable features are
explicitly defined in terms of a few arbitrary parameters and in addition of 
$N$ arbitrary numbers $z_{n}$ or, alternatively, of the $N$ zeros $\bar{z}%
_{n}$ of the polynomials of degree $N$ belonging to the Askey and $q$-Askey
schemes. A terse Section 5 entitled Outlook outlines possible future
developments. Some calculations are confined to an Appendix A in order to
avoid unessential interruptions in the flow of the presentation.

\bigskip

\section{Main results}

Let the two difference operators $\hat{\nabla}\left( a\right) $ and $\check{%
\nabla}\left( q\right) $ be defined as follows: 
\end{subequations}
\begin{equation}
\hat{\nabla}\left( a\right) ~f\left( z\right) =\frac{f\left( z+a\right)
-f\left( z\right) }{a}~,  \label{OpNablaHat}
\end{equation}

\begin{equation}
\check{\nabla}\left( q\right) ~f\left( z\right) =\frac{f\left( z\right)
-f\left( q~z\right) }{\left( 1-q\right) ~z}~.  \label{OpNablaInvHat}
\end{equation}

It is plain that the first of these difference operators becomes the
standard differential operator $d/dz$ as $a\rightarrow 0$, and likewise the
second as $q\rightarrow 1$:%
\begin{equation}
\hat{\nabla}\left( 0\right) ~f\left( z\right) =\check{\nabla}\left( 1\right)
~f\left( z\right) =f^{\prime }\left( z\right) ~.
\end{equation}%
But hereafter we assume for simplicity that $a\neq 0$ and $q\neq 1$.

It is also plain that for the difference operator $\hat{\nabla}\left(
a\right) $ there holds the following eigenvalue equation: 
\begin{subequations}
\label{EigenEqNablaHat}
\begin{equation}
z~\hat{\nabla}\left( a\right) ~\hat{f}_{k}\left( z;a\right) =k~\hat{f}%
_{k}\left( z;a\right) ~,~~~k=0,~1,~2,...~,  \label{EigenDeltaHat}
\end{equation}%
with the eigenfunction $\hat{f}_{k}\left( z;a\right) $ coinciding---up to an
irrelevant multiplicative constant---with the "shifted-factorial" $\left(
z,a\right) _{k}$,%
\begin{equation}
\hat{f}_{k}\left( z;a\right) =\left( z,a\right) _{k}~,
\end{equation}%
itself defined (here and hereafter) as follows:%
\begin{equation}
\left( z,a\right) _{0}=1~;~~~\left( z,a\right)
_{r}=\dprod\limits_{s=0}^{r-1}\left( z+s~a\right) ~,~~~r=1,2,...~.
\label{Pochhammer}
\end{equation}%
Likewise for the difference operator $\check{\nabla}\left( q\right) $ there
holds the following eigenvalue equation: 
\end{subequations}
\begin{subequations}
\begin{equation}
z~\check{\nabla}\left( q\right) ~\check{f}_{k}\left( z\right) =\frac{1-q^{k}%
}{1-q}~\check{f}_{k}\left( z\right) ~,~~~\check{f}_{k}\left( z\right)
=z^{k}~.  \label{EigenEqNablaInvHata}
\end{equation}

\textit{Remark 2.1}. In the eigenvalue equation (\ref{EigenEqNablaHat}) the 
\textit{nonnegative integer} eigenvalues $k$ are independent of the
parameter $a$, while the corresponding eigenfunctions $\hat{f}_{k}\left(
z;a\right) =\left( z;a\right) _{k}$ depend on both $k$ and $a$. Viceversa,
in the eigenvalue equation (\ref{EigenEqNablaInvHata}) the eigenfunctions $%
\check{f}_{k}\left( z\right) =z^{k}$ do not depend on the parameter $q,$
while the corresponding eigenvalues $\left( 1-q^{k}\right) /\left(
1-q\right) $ depend on both $k$ and the parameter $q$; in this case $k$ is 
\textit{a priori} not restricted to take \textit{integer}, or even \textit{%
real}, values, but in our treatment we will in fact restrict attention also
in this case to \textit{nonnegative integer} values of the parameter $k$, so
that the eigenfunction $\check{f}_{k}\left( z\right) =z^{k}$ is holomorphic,
and the eigenvalues in (\ref{EigenEqNablaInvHata}) read as follows:%
\begin{eqnarray}
\frac{1-q^{k}}{1-q} &=&0~~~\text{for \ }k=0~,  \notag \\
\frac{1-q^{k}}{1-q} &=&1+q+q^{2}+...+q^{k-1}=\sum_{s=0}^{k-1}\left(
q^{s}\right) ~~\text{for~~}k=1,2,3,...~\square ~.
\end{eqnarray}

\textit{Notation 2.1}. It is convenient to also introduce the simpler
operators $\hat{\delta}\left( a\right) $ and $\check{\delta}\left( q\right) $
defined as follows: 
\end{subequations}
\begin{subequations}
\label{OpNabladeltaHat}
\begin{equation}
\hat{\delta}\left( a\right) ~f\left( z\right) =f\left( z+a\right) ~,~\ \ 
\check{\delta}\left( q\right) ~f\left( z\right) =f\left( q~z\right) ~.
\label{OpdeltaHat}
\end{equation}%
This implies of course that the operators $\hat{\nabla}\left( a\right) $ and 
$\check{\nabla}\left( q\right) $ defined above are related to these
operators---acting on functions $f\left( z\right) $ of the variable $z$---as
follows:%
\begin{equation}
\hat{\nabla}\left( a\right) =a^{-1}~\left[ \hat{\delta}\left( a\right) -1%
\right] ~,~~~\check{\nabla}\left( q\right) =\left[ \left( 1-q\right) ~z%
\right] ^{-1}~\left[ 1-\check{\delta}\left( q\right) \right] ~.~\square
\label{NablaDelta}
\end{equation}

The main result of this paper is the identification of two $\left( N\times
N\right) $-matrices $\underline{\underline{\hat{\delta}}}\left( a;\underline{%
z}\right) $ and $\underline{\underline{\check{\delta}}}\left( q;\underline{z}%
\right) $ which provide---together with the $N\times N$ diagonal matrix $%
\underline{\underline{Z}}=$diag$\left[ z_{n}\right] ,$ see (\ref{MatZ}%
)---finite-dimensional representations of the two operators $\hat{\delta}%
\left( a\right) $ respectively $\check{\delta}\left( q\right) $. These
matrices are defined componentwise as follows: 
\end{subequations}
\begin{subequations}
\label{Matdeltas}
\begin{equation}
\left( \underline{\underline{\hat{\delta}}}\left( a;\underline{z}\right)
\right) _{nm}=\prod\limits_{\ell =1,~\ell \neq m}^{N}\left( \frac{%
z_{n}+a-z_{\ell }}{z_{m}-z_{\ell }}\right) ~;  \label{MatdeltaHat}
\end{equation}%
\begin{equation}
\left( \underline{\underline{\check{\delta}}}\left( q;\underline{z}\right)
\right) _{nm}=\prod\limits_{\ell =1,~\ell \neq m}^{N}\left( \frac{%
q~z_{n}-z_{\ell }}{z_{m}-z_{\ell }}\right) ~.  \label{MatdeltaInvHat}
\end{equation}%
And of course the $\left( N\times N\right) $-matrices $\underline{\underline{%
\hat{\nabla}}}\left( a;\underline{z}\right) =\left[ \underline{\underline{%
\hat{\delta}}}\left( a;\underline{z}\right) -1\right] /a$ respectively $%
\underline{\underline{\check{\nabla}}}\left( q;\underline{z}\right) =\left[
\left( 1-q\right) ~\underline{\underline{Z}}\left( \underline{z}\right) %
\right] ^{-1}\left[ 1-\underline{\underline{\check{\delta}}}\left( q;%
\underline{z}\right) \right] $ (see (\ref{NablaDelta}) and (\ref{Matdeltas}%
)) which provide---again, together with the $N\times N$ diagonal matrix $%
\underline{\underline{Z}}$---finite-dimensional representations of the two
difference operators $\hat{\nabla}\left( a\right) $ respectively $\check{%
\nabla}\left( q\right) $ are correspondingly defined componentwise as
follows: 
\end{subequations}
\begin{subequations}
\label{MatDeltas}
\begin{equation}
\left( \underline{\underline{\hat{\nabla}}}\left( a;\underline{z}\right)
\right) _{nm}=a^{-1}~\left\{ \left[ \prod\limits_{\ell =1,~\ell \neq
m}^{N}\left( \frac{z_{n}+a-z_{\ell }}{z_{m}-z_{\ell }}\right) \right]
-\delta _{nm}\right\} ~;  \label{MatNablaHat}
\end{equation}%
\begin{equation}
\left( \underline{\underline{\check{\nabla}}}\left( q;\underline{z}\right)
\right) _{nm}=\left[ \left( 1-q\right) ~z_{n}\right] ^{-1}~\left\{ \left[
\prod\limits_{\ell =1,~\ell \neq m}^{N}\left( \frac{z_{n}+a-z_{\ell }}{%
z_{m}-z_{\ell }}\right) \right] -\delta _{nm}\right\} ~.
\label{MatNablaInvHat}
\end{equation}

\textit{Notation 2.2}. In all the above formulas and hereafter $\delta _{nm}$
is the standard Kronecker symbol, 
\end{subequations}
\begin{equation}
\delta _{nm}=1~~~\text{if \ \ }n=m~,~~~\delta _{nm}=0~~~\text{if \ \ }n\neq
m~;  \label{Kronecker}
\end{equation}%
while---unless otherwise explicitly specified---the $N$ numbers $z_{n}$ are 
\textit{arbitrary} (but obviously all different among themselves, and the 
\textit{same} set of $N$ numbers throughout). Let us reiterate that all the $%
\left( N\times N\right) $-matrices introduced above are defined in terms of
these $N$ \textit{a priori} arbitrary numbers, i. e. of the $N$ components
of the $N$-vector $\underline{z}$. This important fact should always be kept
in mind, even though, for notational simplicity, we occasionally omit below
to indicate explicitly this dependence. $\square $

The significance of the statement that the $\left( N\times N\right) $%
-matrices $\underline{\underline{\hat{\delta}}}\left( a;\underline{z}\right) 
$, $\underline{\underline{\check{\delta}}}\left( q;\underline{z}\right) $, $%
\underline{\underline{\hat{\nabla}}}\left( a;\underline{z}\right) $
respectively $\underline{\underline{\check{\nabla}}}\left( q;\underline{z}%
\right) $ provide finite-dimensional representations of the corresponding
operators $\hat{\delta}\left( a\right) $, $\check{\delta}\left( q\right) $, $%
\hat{\nabla}\left( a\right) $ respectively $\check{\nabla}\left( q\right) $%
---and, most importantly, that these representations are \textit{exact} in
the functional space spanned by polynomials of degree less than $N$---is
made explicit by the following properties which extend to these operators
results reported in Section 2.4 of \cite{C2001} for \textit{differential}
operators (as tersely summarized in the preceding Section 1). The main idea
is again to identify relations among operators acting on functions $f\left(
z\right) $---always restricted to live in the functional space spanned by
polynomials in $z$ of degree less than $N$, see (\ref{Polf})---and
appropriately defined matrices acting on the corresponding $N$-vector $%
\underline{f}$, see (\ref{MatZ}), (\ref{Matdeltas}) and (\ref{MatDeltas}).

We report in the following Sections 2.1 respectively 2.2 the main relevant
formulas for the operators $\hat{\delta}\left( a\right) $ and $\hat{\Delta}%
\left( a\right) $ respectively $\check{\delta}\left( q\right) $ and $\check{%
\Delta}\left( q\right) $; proofs of those of these results that are not
immediately obvious are postponed to Section 4. And in Appendix B we report
a few additional \textit{remarkable} properties of the two $\left( N\times
N\right) $-matrices $\underline{\underline{\hat{\delta}}}\left( a;\underline{%
z}\right) $ and $\underline{\underline{\check{\delta}}}\left( q;\underline{z}%
\right) $.

\bigskip

\subsection{The $\left( N\times N\right) $-matrices $\protect\underline{%
\protect\underline{\hat{\protect\delta}}}\left( a\right) $ and $\protect%
\underline{\protect\underline{\hat{\protect\nabla}}}\left( a\right) $}

In this Section 2.1 we report the main results concerning finite-dimensional
representations of the operators $\hat{\delta}\left( a\right) $ and $\hat{%
\nabla}\left( a\right) ,$ see (\ref{OpNablaHat}), (\ref{OpdeltaHat}) and (%
\ref{NablaDelta}).

\textit{Lemma 2.1.1}. Let $f\left( z\right) $ be an arbitrary polynomial in $%
z$ of degree less than $N$, see (\ref{Polf}), and let us denote with the
notation $\hat{f}^{\left[ a,r\right] }\left( z\right) $ the polynomial (also
of degree less than $N$) that obtains by applying to it $r$ times the
operator $\hat{\delta}\left( a\right) $, see (\ref{OpdeltaHat}): 
\begin{subequations}
\begin{eqnarray}
&&\hat{f}^{\left[ a,r\right] }\left( z\right) \equiv \left( \hat{\delta}%
\left( a\right) \right) ^{r}~f\left( z\right) =f\left( z+r~a\right)  \notag
\\
&=&\sum_{m=1}^{N}\left[ c_{m}~\left( z+r~a\right) ^{N-m}\right] =\hat{f}^{%
\left[ ar\right] }\left( z\right) ~,~~~r=1,2,...~.
\end{eqnarray}

Now associate to $f\left( z\right) $ respectively to $\hat{f}^{\left[ ar%
\right] }\left( z\right) $ the $N$-vectors $\underline{f}\equiv \underline{f}%
\left( \underline{z}\right) $ respectively $\underline{\hat{f}}^{\left[ ar%
\right] }\equiv \underline{\hat{f}}^{\left[ ar\right] }\left( \underline{z}%
\right) $, whose $N$ components $f_{n}$ respectively $\hat{f}_{n}^{\left[ ar%
\right] }$ are the $N$ values that the polynomials (of degree less than $N,$
see (\ref{Polf})) $f\left( z\right) $ respectively $\hat{f}^{\left[ ar\right]
}\left( z\right) =f\left( z+r~a\right) $ take at the $N$ (arbitrary) points $%
z_{n}$, 
\begin{eqnarray}
f_{n} &=&f\left( z_{n}\right) ~,~~~\hat{f}_{n}^{\left[ ar\right] }=\hat{f}^{%
\left[ ar\right] }\left( z_{n}\right) =f\left( z_{n}+a~r\right) ~,  \notag \\
n &=&1,...,N~;~~~r=0,1,2,...~.
\end{eqnarray}%
There holds then the $N$-vector formula 
\end{subequations}
\begin{equation}
\underline{\hat{f}}^{\left[ ar\right] }=\left[ \underline{\underline{\hat{%
\delta}}}\left( a;\underline{z}\right) \right] ^{r}~\underline{f}%
~,~~~r=0,1,2,...~,  \label{Prop21}
\end{equation}%
with the $\left( N\times N\right) $-matrices $\underline{\underline{\hat{%
\delta}}}\left( a;\underline{z}\right) $ defined componentwise by (\ref%
{MatdeltaHat}). $\square $

\textit{Remark 2.1.1}. The fact that the matrix $\left[ \underline{%
\underline{\hat{\delta}}}\left( a;\underline{z}\right) \right] ^{r}$
appearing in the right-hand side of the last equation depends---consistently
with the left-hand side of this equation---on the product $ar$ (rather than
separately on $a$ and $r$) is not immediately obvious from its definition (%
\ref{MatdeltaHat}), but is in fact true, indeed see below \textit{Remark
4.1.1}. $\square $

Clearly this finding implies an analogous result for the \textit{difference}
operator $\hat{\nabla}\left( a\right) $, see (\ref{NablaDelta}):

\textit{Lemma 2.1.2}. Let $f\left( z\right) $ be an arbitrary polynomial in $%
z$ of degree less than $N$, see (\ref{Polf}), and let us denote with the
notation $\hat{f}^{\left[ \left[ a,r\right] \right] }\left( z\right) $ the
polynomial that obtains by applying to it $r$ times the difference operator $%
\hat{\nabla}\left( a\right) $, see (\ref{OpNablaHat}) and (\ref{NablaDelta}%
): 
\begin{subequations}
\begin{equation}
\hat{f}^{\left[ \left[ a,r\right] \right] }\left( z\right) \equiv \left( 
\hat{\nabla}\left( a\right) \right) ^{r}~f\left( z\right) =\left[ \frac{\hat{%
\delta}\left( a\right) -1}{a}\right] ^{r}~f\left( z\right) ~,~~~r=0,1,2,...~.
\label{fHatar}
\end{equation}%
Now associate to $f\left( z\right) $ respectively to $\hat{f}^{\left[ \left[
a,r\right] \right] }\left( z\right) $ the $N$-vectors $\underline{f}\equiv 
\underline{f}\left( \underline{z}\right) $ respectively $\underline{\hat{f}}%
^{\left[ \left[ a,r\right] \right] }\equiv \underline{\hat{f}}^{\left[ \left[
a,r\right] \right] }\left( \underline{z}\right) $, whose $N$ components $%
f_{n}\equiv f_{n}\left( \underline{z}\right) $ respectively $f_{n}^{\left[ %
\left[ a,r\right] \right] }\equiv f_{n}^{\left[ \left[ a,r\right] \right]
}\left( \underline{z}\right) $ are the $N$ values that the polynomials $%
f\left( z\right) $ respectively $f^{\left[ \left[ a,r\right] \right] }\left(
z\right) $ take at the $N$ (arbitrary) points $z_{n}$, 
\begin{equation}
f_{n}=f\left( z_{n}\right) ~,~~~f_{n}^{\left[ \left[ a,r\right] \right] }=f^{%
\left[ \left[ a,r\right] \right] }\left( z_{n}\right) ~,~~~n=1,...,N~.
\end{equation}%
There holds then the $N$-vector formula 
\end{subequations}
\begin{equation}
\underline{\hat{f}}^{\left[ \left[ a,r\right] \right] }=\left[ \underline{%
\underline{\hat{\nabla}}}\left( a;\underline{z}\right) \right] ^{r}~%
\underline{f}~,~~~r=0,1,2,...~,
\end{equation}%
of course with the $\left( N\times N\right) $-matrix $\underline{\underline{%
\hat{\nabla}}}\left( a;\underline{z}\right) $ defined componentwise by (\ref%
{MatNablaHat}). $\square $

\textit{Remark 2.1.2}. It is plain that the operator $\hat{\nabla}\left(
a\right) ,$ when acting on a polynomial in $z$ of degree $M$, yields a
polynomial of degree $M-1$; hence, when it acts $r$ times on any polynomial
of degree less than $N$ it yields an identically vanishing result if the
integer $r$ equals or exceeds $N.$ Hence the right-hand side of (\ref{fHatar}%
) vanishes for $r\geq N,$ and this implies that the $(N\times N)$-matrix $%
\underline{\underline{\hat{\nabla}}}\left( a;\underline{z}\right) $ features
the \textit{remarkable} property%
\begin{equation}
\left[ \underline{\underline{\hat{\nabla}}}\left( a;\underline{z}\right) %
\right] ^{N}=\underline{\underline{0}}~,  \label{DeltaHatZero}
\end{equation}%
where $\underline{\underline{0}}$ denotes of course the $(N\times N)$-matrix
with all elements vanishing. $\square $

The following \textit{Proposition }and \textit{Corollaries} are immediate
consequences of these findings.

\textit{Proposition 2.1.1}. Let the difference operator $\hat{D}\left(
a\right) $ be defined as follows,%
\begin{equation}
\hat{D}\left( a\right) =\sum_{r=0}^{R}\left\{ \hat{d}_{r}\left( z\right) ~%
\left[ \hat{\delta}\left( a\right) \right] ^{r}\right\} ~,  \label{Dhat}
\end{equation}%
where the positive integer $R$ and the $R+1$ functions $\hat{d}_{r}\left(
z\right) $ are \textit{a priori} arbitrary (but see below the restriction on
the function $f\left( z\right) $), and let%
\begin{equation}
\hat{D}\left( a\right) ~f\left( z\right) =g\left( z\right)  \label{Dhatfg}
\end{equation}%
with $f\left( z\right) $ a polynomial in $z$ of degree less than $N,$ see (%
\ref{Polf}) (but note: no such condition on $g\left( z\right) $). There then
holds the $N$-vector equation%
\begin{equation}
\underline{\underline{\hat{D}}}\left( a;\underline{z}\right) ~\underline{f}%
\left( \underline{z}\right) =\underline{g}\left( \underline{z}\right)
\label{DHatfg}
\end{equation}%
with the $\left( N\times N\right) $-matrix $\underline{\underline{\hat{D}}}%
\left( a;\underline{z}\right) $ defined as follows, 
\begin{equation}
\underline{\underline{\hat{D}}}\left( a;\underline{z}\right)
=\sum_{r=0}^{R}\left\{ \hat{d}_{r}\left( \underline{\underline{Z}}\right) ~%
\left[ \underline{\underline{\hat{\delta}}}\left( a;\underline{z}\right) %
\right] ^{r}\right\} ~,  \label{MatDhat}
\end{equation}%
and of course the $N$-vectors $\underline{f}\left( \underline{z}\right) $
and $\underline{g}\left( \underline{z}\right) $ defined so that their $N$
components are 
\begin{subequations}
\label{Componentsfg}
\begin{equation}
\left( \underline{f}\left( \underline{z}\right) \right) _{n}=f\left(
z_{n}\right) ~,~~~\left( \underline{g}\left( \underline{z}\right) \right)
_{n}=g\left( z_{n}\right) ~,~~~n=1,...,N~,  \label{Componentsfga}
\end{equation}%
or, equivalently, 
\begin{equation}
\underline{f}\left( \underline{z}\right) =f\left( \underline{\underline{Z}}%
\right) ~\underline{u}~,~~~\underline{g}\left( \underline{z}\right) =g\left( 
\underline{\underline{Z}}\right) ~\underline{u}~,  \label{Componentsfgb}
\end{equation}%
of course with the $\left( N\times N\right) $-matrices $\underline{%
\underline{\hat{\delta}}}\left( a;\underline{z}\right) $ respectively $%
\underline{\underline{Z}}$ defined componentwise by (\ref{MatdeltaHat})
respectively (\ref{MatZ}) and the "unit" $N$-vector $\underline{u}$ defined
by (\ref{Vecu}). $\square $

\textit{Corollary 2.1.1}. If in (\ref{Dhatfg}) $g\left( z\right) =0,$ i. e.
if for the operator $\hat{D}\left( a\right) $, see (\ref{Dhat}), there holds
the equation 
\end{subequations}
\begin{subequations}
\begin{equation}
\hat{D}\left( a\right) ~f\left( z\right) =0~,
\end{equation}%
with $f\left( z\right) $ a polynomial in $z$ of degree less than $N,$ see (%
\ref{Polf}), then the $\left( N\times N\right) $-matrix $\underline{%
\underline{\hat{D}}}\left( a;\underline{z}\right) $, see (\ref{MatDhat}),
has vanishing determinant,\ 
\begin{equation}
\det \left[ \underline{\underline{\hat{D}}}\left( a;\underline{z}\right) %
\right] =0~.~\square
\end{equation}

\textit{Corollary 2.1.2}. If the operator $\hat{D}\left( a\right) $, see (%
\ref{Dhat}), has the eigenvalue $b$, 
\end{subequations}
\begin{subequations}
\begin{equation}
\hat{D}\left( a\right) ~f^{\left( b\right) }\left( z\right) =b~f^{\left(
b\right) }\left( z\right)
\end{equation}%
with the corresponding eigenfunction $f^{\left( b\right) }\left( z\right) $
being a polynomial in $z$ of degree less than $N$, see (\ref{Polf}), then
the corresponding $\left( N\times N\right) $-matrix $\underline{\underline{%
\hat{D}}}\left( a;\underline{z}\right) $, see (\ref{MatDhat}), features the
same eigenvalue $b$,%
\begin{equation}
\underline{\underline{\hat{D}}}\left( a;\underline{z}\right) ~\underline{f}%
^{\left( b\right) }=b~\underline{f}^{\left( b\right) }~,
\end{equation}%
and the corresponding eigenvector $\underline{f}^{\left( b\right) }$ is
given by the following simple prescription,%
\begin{equation}
\underline{f}^{\left( b\right) }=f^{\left( b\right) }\left( \underline{%
\underline{Z}}\right) ~\underline{u}~;~~~\left( \underline{f}^{\left(
b\right) }\right) _{n}=f^{\left( b\right) }\left( z_{n}\right)
~,~~~n=1,...,N~,
\end{equation}%
where of course the $\left( N\times N\right) $-matrix $\underline{\underline{%
Z}}$, respectively the $N$-vector $\underline{u}$, are again defined by (\ref%
{MatZ}) respectively (\ref{Vecu}). $\square $

Clearly these equations are merely examples of the neat prescriptions 
\end{subequations}
\begin{eqnarray}
\hat{d}_{s}\left( z\right) &\Rightarrow &\hat{d}_{s}\left( \underline{%
\underline{Z}}\left( \underline{z}\right) \right) ~,~~~\hat{\delta}\left(
a\right) \Rightarrow \underline{\underline{\hat{\delta}}}\left( a;\underline{%
z}\right) ~,~~~\hat{\nabla}\left( a\right) \Rightarrow \underline{\underline{%
\hat{\nabla}}}\left( a;\underline{z}\right) ~,  \notag \\
f\left( z\right) &\Rightarrow &\underline{f}\left( \underline{z}\right)
=f\left( \underline{\underline{Z}}\right) ~\underline{u}~,
\end{eqnarray}%
which allow to transform equations involving the action of the
multiplicative operator $\hat{d}_{s}\left( z\right) $ (see (\ref{Dhat})) and
of the operators $\hat{\delta}\left( a\right) $ and $\hat{\nabla}\left(
a\right) $ (see (\ref{OpdeltaHat}) and (\ref{OpNablaHat}) or (\ref%
{NablaDelta})) acting on functions $f\left( z\right) $, into corresponding
equations involving the action of corresponding $\left( N\times N\right) $%
-matrices on corresponding $N$-vectors; rules that involve the introduction
of $N$ \textit{arbitrary} numbers $z_{n}$ (all different among themselves),
and that are applicable whenever these operators act on functions $f\left(
z\right) $ which are polynomials in $z$ of degree less than the \textit{%
arbitrary} number $N$, and that involve the simultaneous replacement of the
function $f\left( z\right) $ into the $N$-vector $\underline{f}\left( 
\underline{z}\right) $ of components $f_{n}\left( \underline{z}\right)
=f\left( z_{n}\right) $.

\textit{Remark 2.1.3}. An interesting generalization of all the findings
reported above (in this Section 2.1) is to the case in which the function $%
f\left( z\right) $, instead of being a polynomial of degree less than $N$ in 
$z$, is a polynomial of degree less than $N$ in a variable $\zeta \equiv
\zeta \left( z\right) $, 
\begin{equation}
f\left( \zeta \right) \equiv f\left( \zeta \left( z\right) \right)
=\sum_{m=1}^{N}\left\{ c_{m}~\left[ \zeta \left( z\right) \right]
^{N-m}\right\} ~.  \label{fzita}
\end{equation}%
It is then easily seen that all the results reported above (in this Section
2.1) remain valid, provided the $\left( N\times N\right) $-matrix $%
\underline{\underline{\hat{\delta}}}\left( a;\underline{z}\right) $, see (%
\ref{MatdeltaHat}), is replaced by the $\left( N\times N\right) $-matrix $%
\underline{\underline{\tilde{\delta}}}\left( a;\underline{z}\right) $
defined componentwise as follows: 
\begin{subequations}
\begin{equation}
\left( \underline{\underline{\tilde{\delta}}}\left( a;\underline{z}\right)
\right) _{nm}=\prod\limits_{\ell =1,~\ell \neq m}^{N}\left[ \frac{\zeta
\left( z_{n}+a\right) -\zeta _{\ell }}{\zeta _{m}-\zeta _{\ell }}\right]
~,~\ ~n,m=1,...,N~,
\end{equation}%
where of course 
\begin{equation}
\zeta _{n}\equiv \zeta \left( z_{n}\right) ~,~~~\underline{\zeta }=\left(
\zeta _{1},...,\zeta _{N}\right) ~.
\end{equation}%
Of course likewise the matrix $\underline{\underline{\hat{\nabla}}}\left( a;%
\underline{z}\right) $ is replaced by the matrix $\underline{\underline{%
\tilde{\nabla}}}\left( a;\underline{z}\right) =\left[ \underline{\underline{%
\tilde{\delta}}}\left( a;\underline{z}\right) -1\right] /a$, of components 
\end{subequations}
\begin{eqnarray}
\left( \underline{\underline{\tilde{\nabla}}}\left( a;\underline{z}\right)
\right) _{nm} &=&a^{-1}~\left\{ \prod\limits_{\ell =1,~\ell \neq m}^{N}\left[
\frac{\zeta \left( z_{n}+a\right) -\zeta _{\ell }}{\zeta _{m}-\zeta _{\ell }}%
\right] -\delta _{nm}\right\} ~,  \notag \\
n,m &=&1,...,N~;
\end{eqnarray}%
and the $N$-vector $\underline{f}\left( \underline{z}\right) $ of components 
$f_{n}\left( \underline{z}\right) =f\left( z_{n}\right) $ is replaced by the 
$N$-vector $\underline{f}\left( \underline{\zeta }\left( \underline{z}%
\right) \right) $ of components $f_{n}\left( \underline{\zeta }\left( 
\underline{z}\right) \right) =f\left( \zeta _{n}\right) ,$ while the $N$%
-vector $\underline{f}\left( \underline{z}+a~\underline{u}\right) $ of
components $f_{n}\left( \underline{z}+a~\underline{u}\right) =f\left(
z_{n}+a\right) $ is replaced by the $N$-vector $\underline{f}\left( 
\underline{\zeta }\left( \underline{z}+a~\underline{u}\right) \right) $ of
components $f_{n}\left( \underline{z}+a~\underline{u}\right) =f\left( \zeta
\left( z_{n}+a\right) \right) $. $\square $

The proof of this \textit{Remark 2.1.3 }is quite analogous to that of 
\textit{Lemma 2.1.1} (see Section 4) and is therefore omitted.

\bigskip

\subsection{The $\left( N\times N\right) $-matrices $\protect\underline{%
\protect\underline{\check{\protect\delta}}}\left( q;\protect\underline{z}%
\right) $ and $\protect\underline{\protect\underline{\check{\protect\nabla}}}%
\left( q;\protect\underline{z}\right) $}

In this Section 2.2 we report the main results concerning finite-dimensional
representations of the operators $\check{\delta}\left( q\right) $ and $%
\check{\nabla}\left( q\right) ,$ see (\ref{OpNablaInvHat}), (\ref{OpdeltaHat}%
) and (\ref{NablaDelta}). It should be mentioned that a somewhat less
explicit, but essentially equivalent, finite-dimensional representation of
the difference operator $\check{\nabla}\left( q\right) $ was already
provided almost two decades ago by Chakrabarti and Jagannathan \cite{CJ1995}%
; and also, in the context of a more special case ("3-body problem") in \cite%
{CX1995}.

\textit{Lemma 2.2.1}. Let $f\left( z\right) $ be an arbitrary polynomial in $%
z$ of degree less than $N$, see (\ref{Polf}), and let us denote with the
notation $\check{f}^{\left\{ q,r\right\} }\left( z\right) $ the polynomial
(clearly of the same degree) that obtains by applying to it $r$ times the
operator $\check{\delta}\left( q\right) $, see (\ref{OpdeltaHat}): 
\begin{subequations}
\begin{eqnarray}
\check{f}^{\left\{ q,r\right\} }\left( z\right) &\equiv &\left( \check{\delta%
}\left( q\right) \right) ^{r}~f\left( z\right) =f\left( q^{r}~z\right)
=\sum_{m=1}^{N}\left[ c_{m}~\left( q^{r}~z\right) ^{N-m}\right] =\check{f}%
^{\left\{ q^{r}\right\} }\left( z\right) ~,  \notag \\
r &=&1,2,,3...~.
\end{eqnarray}

Now associate to $f\left( z\right) $ respectively to $\check{f}^{\left\{
q^{r}\right\} }\left( z\right) $ the $N$-vectors $\underline{f}\equiv 
\underline{f}\left( \underline{z}\right) $ respectively $\underline{\check{f}%
}^{\left\{ q^{r}\right\} }\equiv \underline{\check{f}}^{\left\{
q^{r}\right\} }\left( \underline{z}\right) $, whose $N$ components $f_{n}$
respectively $\check{f}_{n}^{\left\{ q^{r}\right\} }$ are the $N$ values
that the polynomials $f\left( z\right) $ respectively $\check{f}^{\left\{
q^{r}\right\} }\left( z\right) =f\left( q^{r}~z\right) $ take at the $N$
(arbitrary) points $z_{n}$, 
\begin{eqnarray}
f_{n} &=&f\left( z_{n}\right) ~,~~~\check{f}_{n}^{\left\{ q^{r}\right\} }=%
\check{f}^{\left\{ q^{r}\right\} }\left( z_{n}\right) =f\left(
q^{r}~z_{n}\right) ~,  \notag \\
n &=&1,...,N~;~~~r=0,1,2,...~.
\end{eqnarray}%
There holds then the $N$-vector formula 
\end{subequations}
\begin{equation}
\underline{\check{f}}^{\left\{ q^{r}\right\} }=\left[ \underline{\underline{%
\check{\delta}}}\left( q;\underline{z}\right) \right] ^{r}~\underline{f}%
~,~~~r=0,1,2,...~,
\end{equation}%
with the $\left( N\times N\right) $-matrices $\underline{\underline{\check{%
\delta}}}\left( q;\underline{z}\right) $ defined componentwise by (\ref%
{MatdeltaInvHat}). $\square $

\textit{Remark 2.2.1}. The fact that the matrix $\left[ \underline{%
\underline{\check{\delta}}}\left( q;\underline{z}\right) \right] ^{r}$
appearing in the right-hand side of the last equation depends---consistently
with the left-hand side of this equation---on the single quantity $q^{r}$
(rather than separately on $q$ and $r$) is not immediately obvious from its
definition (\ref{MatdeltaInvHat}), but is in fact true, indeed see below 
\textit{Remark 4.2.1}. $\square $

Clearly this finding implies an analogous result for the \textit{difference}
operator $\check{\nabla}\left( q\right) $, see (\ref{NablaDelta}):

\textit{Lemma 2.2.2}. Let $f\left( z\right) $ be an arbitrary polynomial in $%
z$ of degree less than $N$, see (\ref{Polf}), and let us denote with the
notation $\check{f}^{\left\{ \left\{ q,r\right\} \right\} }\left( z\right) $
the polynomial that obtains by applying to it $r$ times the difference
operator $\check{\nabla}\left( q\right) $, see (\ref{OpNablaInvHat}) and (%
\ref{NablaDelta}): 
\begin{subequations}
\begin{equation}
\check{f}^{\left\{ \left\{ q,r\right\} \right\} }\left( z\right) \equiv
\left( \check{\nabla}\left( q\right) \right) ^{r}~f\left( z\right) =\left[ 
\frac{\check{\delta}\left( q\right) -1}{\left( q-1\right) ~z}\right]
^{r}~f\left( z\right) ~,~~~r=0,1,2,...~.  \label{fInvHatar}
\end{equation}%
Now associate to $f\left( z\right) $ respectively to $\check{f}^{\left\{
\left\{ q,r\right\} \right\} }\left( z\right) $ the $N$-vectors $\underline{f%
}\equiv \underline{f}\left( \underline{z}\right) $ respectively $\underline{%
\check{f}}^{\left\{ q^{r}\right\} }\equiv \underline{\check{f}}^{\left\{
q^{r}\right\} }\left( \underline{z}\right) $, whose $N$ components $%
f_{n}\equiv f_{n}\left( \underline{z}\right) $ respectively $\check{f}%
_{n}^{\left\{ \left\{ q,r\right\} \right\} }\equiv \check{f}_{n}^{\left\{
\left\{ q,r\right\} \right\} }\left( \underline{z}\right) $ are the $N$
values that the polynomials $f\left( z\right) $ respectively $\check{f}%
^{\left\{ \left\{ q,r\right\} \right\} }\left( z\right) $ take at the $N$
(arbitrary) points $z_{n}$, 
\begin{equation}
f_{n}=f\left( z_{n}\right) ~,~~~\check{f}_{n}^{\left\{ \left\{ q,r\right\}
\right\} }=\check{f}^{\left\{ \left\{ q,r\right\} \right\} }\left(
z_{n}\right) ~,~~~n=1,...,N~.
\end{equation}%
There holds then the $N$-vector formula 
\end{subequations}
\begin{equation}
\underline{\check{f}}^{\left\{ q^{r}\right\} }=\left[ \underline{\underline{%
\check{\nabla}}}\left( q;\underline{z}\right) \right] ^{r}~\underline{f}%
~,~~~r=0,1,2,...~,
\end{equation}%
of course with the $\left( N\times N\right) $-matrix $\underline{\underline{%
\check{\nabla}}}\left( q;\underline{z}\right) $ defined componentwise by (%
\ref{MatNablaInvHat}). $\square $

\textit{Remark 2.2.2}. It is again plain that the operator $\check{\nabla}%
\left( q\right) ,$ when acting on a polynomial in $z$ of degree $M$, yields
a polynomial of degree $M-1$; hence, when it acts $r$ times on any
polynomial of degree less than $N$ it yields an identically vanishing result
if the integer $r$ equals or exceeds $N.$ Hence the right-hand side of (\ref%
{fInvHatar}) vanishes for $r\geq N,$ and this implies that the $(N\times N)$%
-matrix $\underline{\underline{\check{\nabla}}}\left( q;\underline{z}\right) 
$ features the \textit{remarkable} property%
\begin{equation}
\left[ \underline{\underline{\check{\nabla}}}\left( q;\underline{z}\right) %
\right] ^{N}=\underline{\underline{0}}~,  \label{DeltaInvHatZero}
\end{equation}%
where again $\underline{\underline{0}}$ denotes the $(N\times N)$-matrix
with all elements vanishing. $\square $

The following \textit{Proposition }and \textit{Corollaries} are immediate
consequences of these findings.

\textit{Proposition 2.2.1}. Let the difference operator $\check{D}\left(
q\right) $ be defined as follows,%
\begin{equation}
\check{D}\left( q\right) =\sum_{r=0}^{R}\left\{ \check{d}_{r}\left( z\right)
~\left[ \check{\delta}\left( q\right) \right] ^{r}\right\} ~,
\end{equation}%
where the positive integer $R$ and the $R+1$ functions $\check{d}_{r}\left(
z\right) $ are \textit{a priori} arbitrary (but see below the restriction on
the function $f\left( z\right) $), and let%
\begin{equation}
\check{D}\left( q\right) ~f\left( z\right) =g\left( z\right)
\label{DInvhatfg}
\end{equation}%
with $f\left( z\right) $ a polynomial in $z$ of degree less than $N,$ see (%
\ref{Polf}) (but note: no such condition on $g\left( z\right) $). There then
holds the $N$-vector equation%
\begin{equation}
\underline{\underline{\check{D}}}\left( q;\underline{z}\right) ~\underline{f}%
\left( \underline{z}\right) =\underline{g}\left( \underline{z}\right)
\label{DInvHatfg}
\end{equation}%
with the $\left( N\times N\right) $-matrix $\underline{\underline{\check{D}}}%
\left( q;\underline{z}\right) $ defined as follows, 
\begin{equation}
\underline{\underline{\check{D}}}\left( q;\underline{z}\right)
=\sum_{r=0}^{R}\left\{ \check{d}_{r}\left( \underline{\underline{Z}}\right) ~%
\left[ \underline{\underline{\check{\delta}}}\left( q;\underline{z}\right) %
\right] ^{r}\right\} ~,  \label{MatDInvHat}
\end{equation}%
of course with the $N$-vectors $\underline{f}\left( \underline{z}\right) $
and $\underline{g}\left( \underline{z}\right) $ defined as above, see (\ref%
{Componentsfg}), and the $\left( N\times N\right) $-matrices $\underline{%
\underline{\check{\delta}}}\left( q;\underline{z}\right) $ respectively $%
\underline{\underline{Z}}$ defined componentwise by (\ref{MatdeltaInvHat})
respectively (\ref{MatZ}). $\square $

\textit{Corollary 2.2.1}. If in (\ref{DInvhatfg}) $g\left( z\right) =0,$ i.
e. if for the operator $\check{D}\left( q\right) $, see (\ref{Dhat}), there
holds the equation~ 
\begin{subequations}
\begin{equation}
\check{D}\left( q\right) ~\underline{f}\left( z\right) =0~,
\end{equation}%
with $f\left( z\right) $ a polynomial in $z$ of degree less than $N,$ see (%
\ref{Polf}), then the $\left( N\times N\right) $-matrix $\underline{%
\underline{\check{D}}}\left( q;\underline{z}\right) $, see (\ref{MatDInvHat}%
), has vanishing determinant,%
\begin{equation}
\det \left[ \underline{\underline{\check{D}}}\left( q;\underline{z}\right) %
\right] =0~.~\square
\end{equation}

\textit{Corollary 2.2.2}. If the operator $\check{D}\left( q\right) $, see (%
\ref{MatDInvHat}), has the eigenvalue $b$, 
\end{subequations}
\begin{subequations}
\begin{equation}
\check{D}\left( q\right) ~\check{f}^{\left( b\right) }\left( z\right) =b~%
\check{f}^{\left( b\right) }\left( z\right)
\end{equation}%
with the corresponding eigenfunction $\check{f}^{\left( b\right) }\left(
z\right) $ being a polynomial in $z$ of degree less than $N$, see (\ref{Polf}%
), then the corresponding $\left( N\times N\right) $-matrix $\underline{%
\underline{\check{D}}}\left( q;\underline{z}\right) $, see (\ref{MatDInvHat}%
), features the same eigenvalue $b$,%
\begin{equation}
\underline{\underline{\check{D}}}\left( q;\underline{z}\right) ~\underline{%
\check{f}}^{\left( b\right) }=b~\underline{\check{f}}^{\left( b\right) }~,
\end{equation}%
and the corresponding eigenvector $\underline{\check{f}}^{\left( b\right) }$
is given by the following simple prescription,%
\begin{equation}
\underline{\check{f}}^{\left( b\right) }=\check{f}^{\left( b\right) }\left( 
\underline{\underline{Z}}\right) ~\underline{u}~,~~~\left( \underline{\check{%
f}}^{\left( b\right) }\right) _{n}=\check{f}^{\left( b\right) }\left(
z_{n}\right) ~,
\end{equation}%
where of course the $\left( N\times N\right) $-matrix $\underline{\underline{%
Z}}$, respectively the $N$-vector $\underline{u}$, are again defined by (\ref%
{MatZ}) respectively (\ref{Vecu}). $\square $

Clearly these equations are merely examples of the neat prescriptions 
\end{subequations}
\begin{eqnarray}
\check{d}_{s}\left( z\right) &\Rightarrow &\check{d}_{s}\left( \underline{%
\underline{Z}}\left( \underline{z}\right) \right) ~,~~~\check{\delta}\left(
q\right) \Rightarrow \underline{\underline{\check{\delta}}}\left( q;%
\underline{z}\right) ~,~~~\check{\nabla}\left( q\right) \Rightarrow 
\underline{\underline{\check{\nabla}}}\left( q;\underline{z}\right) ,  \notag
\\
f\left( z\right) &\Rightarrow &\underline{f}\left( \underline{z}\right)
=f\left( \underline{\underline{Z}}\right) ~\underline{u}
\end{eqnarray}%
which allow to transform equations involving the action of the
multiplicative operator $\check{d}_{s}\left( z\right) $ (see (\ref%
{MatDInvHat})) and of the operators $\check{\delta}\left( q\right) $ and $%
\check{\nabla}\left( q\right) $ (see (\ref{OpdeltaHat}) and (\ref%
{OpNablaInvHat}) or (\ref{NablaDelta})) acting on functions $f\left(
z\right) $, into corresponding equations involving the action of
corresponding $\left( N\times N\right) $-matrices on corresponding $N$%
-vectors; rules that involve the introduction of $N$ \textit{arbitrary}
numbers $z_{n}$ (all different among themselves), and that are applicable
whenever these operators act on functions $f\left( z\right) $ which are
polynomials in $z$ of degree less than the \textit{arbitrary} number $N,$
and that involve the simultaneous replacement of the function $f\left(
z\right) $ into the $N$-vector $\underline{f}\left( \underline{z}\right) $
of components $f_{n}\left( \underline{z}\right) =f\left( z_{n}\right) $.

\textit{Remark 2.2.3}. An interesting generalization of all the findings
reported above (in this Section 2.2) is to the case in which the function $%
f\left( z\right) $, instead of being a polynomial of degree less than $N$ in 
$z$, is a polynomial of degree less than $N$ in a variable $\zeta \equiv
\zeta \left( z\right) ,$ see (\ref{fzita}). It is then easily seen that all
the results reported above (in this Section 2.2) remain valid, provided the $%
\left( N\times N\right) $-matrix $\underline{\underline{\check{\delta}}}%
\left( q;\underline{z}\right) $, see (\ref{MatdeltaHat}), is replaced by the
matrix $\underline{\underline{\breve{\delta}}}\left( q;\underline{z}\right) $
defined componentwise as follows: 
\begin{subequations}
\begin{equation}
\left( \underline{\underline{\breve{\delta}}}\left( q;\underline{z}\right)
\right) _{nm}=\prod\limits_{\ell =1,~\ell \neq m}^{N}\left[ \frac{\zeta
\left( q~z_{n}\right) -\zeta _{\ell }}{\zeta _{m}-\zeta _{\ell }}\right]
~,~\ ~n,m=1,...,N~,
\end{equation}%
where of course 
\begin{equation}
\zeta _{n}\equiv \zeta \left( z_{n}\right) ~,~~~\underline{\zeta }=\left(
\zeta _{1},...,\zeta _{N}\right) ~.
\end{equation}%
Of course likewise the matrix $\underline{\underline{\check{\nabla}}}\left(
q;\underline{z}\right) $ is replaced by the matrix $\underline{\underline{%
\breve{\nabla}}}\left( q;\underline{z}\right) $ of components 
\end{subequations}
\begin{eqnarray}
\left( \underline{\underline{\breve{\nabla}}}\left( q;\underline{z}\right)
\right) _{nm} &=&\left[ \left( q-1\right) ~\zeta _{n}\right] ^{-1}~\left\{
\prod\limits_{\ell =1,~\ell \neq m}^{N}\left[ \frac{\zeta \left(
q~z_{n}\right) -\zeta _{\ell }}{\zeta _{m}-\zeta _{\ell }}\right] -\delta
_{nm}\right\} ~,  \notag \\
n,m &=&1,...,N~;
\end{eqnarray}%
and the $N$-vector $\underline{f}\left( \underline{z}\right) $ of components 
$f_{n}\left( \underline{z}\right) =f\left( z_{n}\right) $ is replaced by the 
$N$-vector $\underline{f}\left( \underline{\zeta }\left( \underline{z}%
\right) \right) $ of components $f_{n}\left( \underline{\zeta }\left( 
\underline{z}\right) \right) =f\left( \zeta _{n}\right) ,$ while the $N$%
-vector $\underline{f}\left( q~\underline{z}\right) $ of components $%
f_{n}\left( q~z_{n}\right) $ is replaced by the $N$-vector $\underline{f}%
\left( \underline{\zeta }\left( q~\underline{z}\right) \right) $ of
components $f_{n}\left( \zeta \left( q~z_{n}\right) \right) $. $\square $

The proof of this \textit{Remark 2.2.3} is quite analogous to that of 
\textit{Lemma 2.2.1} (see Section 4) and is therefore omitted.

\bigskip

\section{Remarkable matrices}

In this Section 3---as examples of applications of the findings reported in
the preceding Section 2---we identify several $\left( N\times N\right) $%
-matrices which are \textit{remarkable} because of some nontrivial
properties they feature, and we report some properties of the zeros of the
polynomials belonging to the Askey and $q$-Askey schemes.

\textit{Proposition 3.1}. The $\left( N\times N\right) $-matrix 
\begin{subequations}
\begin{equation}
\underline{\underline{\hat{K}}}\left( a;\underline{z}\right) =\underline{%
\underline{Z}}~\underline{\underline{\hat{\nabla}}}\left( a;\underline{z}%
\right) ~,
\end{equation}%
hence defined componentwise as follows in terms of the $N+1$ arbitrary
numbers $z_{n}$ and $a$ (see (\ref{MatZ}) and (\ref{MatNablaHat})), 
\begin{equation}
\hat{K}_{nm}\left( a;\underline{z}\right) =\left( \frac{z_{n}}{a}\right)
~\left\{ \left[ \prod\limits_{\ell =1,~\ell \neq m}^{N}\left( \frac{%
z_{n}+a-z_{\ell }}{z_{m}-z_{\ell }}\right) \right] -\delta _{nm}\right\} ~,
\label{MatKHat}
\end{equation}%
features the $N$ nonnegative integers less than $N$ as its $N$ eigenvalues:%
\begin{equation}
\underline{\underline{\hat{K}}}\left( a;\underline{z}\right) ~\underline{v}%
^{\left( k\right) }\left( a;\underline{z}\right) =k~\underline{\hat{v}}%
^{\left( k\right) }\left( a;\underline{z}\right) ~,~~~k=0,1,...,N-1~,
\end{equation}%
\begin{equation}
\hat{v}_{n}^{\left( k\right) }\left( a;\underline{z}\right) =\left(
z_{n};a\right) _{k}=\dprod\nolimits_{s=0}^{k-1}\left( z_{n}+s~a\right) ~,~%
\hspace{0in}~k=0,1,2,...,N-1~.~\square
\end{equation}

This result is an immediate consequence of \textit{Corollary 2.1.2},
together with the eigenvalue equation (\ref{EigenEqNablaHat}) and of course
the definitions (\ref{MatZ}) respectively (\ref{MatNablaHat}) of the $\left(
N\times N\right) $-matrices $\underline{\underline{Z}}$ respectively $%
\underline{\underline{\hat{\nabla}}}\left( a;\underline{z}\right) $ and (\ref%
{Pochhammer}) of the symbol $\left( z;a\right) _{k}$.

\textit{Remark 3.1}. Note the \textit{Diophantine} character of this result,
and the fact that it implies that the $\left( N\times N\right) $-matrix $%
\underline{\underline{\hat{K}}}\left( a;\underline{z}\right) $, which
clearly depends nontrivially on the $N+1$ numbers $z_{n}$ and $a$, see (\ref%
{MatKHat}), is \textit{isospectral} for any variation of these $N+1$
parameters. $\square $

\textit{Proposition 3.2}. The $\left( N\times N\right) $-matrix 
\end{subequations}
\begin{subequations}
\begin{equation}
\underline{\underline{\hat{F}}}\equiv \underline{\underline{\hat{F}}}\left(
\alpha ,c;\underline{z}\right) =\left( c-\underline{\underline{Z}}\right) ~%
\underline{\underline{\hat{\delta}}}\left( -1;\underline{z}\right) +\left(
2-\alpha \right) ~\underline{\underline{Z}}+\left( \alpha -1\right) 
\underline{\underline{Z}}~\underline{\underline{\hat{\delta}}}\left( 1;%
\underline{z}\right) ~,
\end{equation}%
hence defined componentwise as follows in terms of the $N+2$ arbitrary
numbers $z_{n}$ and $\alpha ,c$ (see (\ref{MatZ}) and (\ref{MatdeltaHat})),%
\begin{eqnarray}
\hat{F}_{nm}\left( \alpha ,c;\underline{z}\right) =\left( 2-\alpha \right)
~z_{n}~\delta _{nm}+\left( c-z_{n}\right) ~\prod\limits_{\ell =1,~\ell \neq
m}^{N}\left( \frac{z_{n}-1-z_{\ell }}{z_{m}-z_{\ell }}\right) &&  \notag \\
+\left( \alpha -1\right) ~z_{n}~\prod\limits_{\ell =1,~\ell \neq
m}^{N}\left( \frac{z_{n}+1-z_{\ell }}{z_{m}-z_{\ell }}\right)
~,~~~n,m=1,...,N~, &&
\end{eqnarray}%
has the $N$ eigenvalues $c+\alpha ~k$ with $k$ the first $N$ nonnegative
integers:%
\begin{equation}
\underline{\underline{\hat{F}}}\left( \alpha ,c;\underline{z}\right) ~%
\underline{f}\left( \underline{z},-k;~c;~\alpha \right) =\left( c+\alpha
~k\right) ~\underline{f}\left( \underline{z},-k;~c;~\alpha \right)
~,~~~k=0,1,...,N-1~,
\end{equation}%
the corresponding $N$ eigenvectors~$\underline{f}\left( \underline{z}%
,-k;~c;~\alpha \right) $ being defined componentwise as follows,%
\begin{equation}
f_{n}\left( \underline{z},-k;~c;~\alpha \right) =F\left( z_{n},-k;~c;~\alpha
\right) ~,
\end{equation}%
where $F\left( a,b;c;z\right) $ is the standard hypergeometric function (see
for instance \cite{E1953}),%
\begin{equation}
F\left( a,b;c;z\right) =\sum_{r=0}^{\infty }\left[ \frac{\left( a\right)
_{r}~\left( b\right) _{r}~z^{r}}{r!~\left( c\right) _{r}}\right]
\label{Hyper}
\end{equation}%
(here of course the Pochhammer symbol $\left( x\right) _{r}\equiv \left(
x;1\right) _{r}$ (see (\ref{Pochhammer})) is defined as follows:%
\begin{equation}
\left( x\right) _{0}=1~;~~~\left( x\right) _{r}=x~\left( x+1\right) \cdot
\cdot \cdot \left( x+r-1\right) ~,~~~r=1,2,...~;  \label{Poch}
\end{equation}%
hence for $b=-k$---and generic values of the parameters $c$ and $\alpha $,
as we generally assume---the sum in the right-hand side of (\ref{Hyper})
stops at $r=k,$ so that $F\left( z,-k;c;\alpha \right) $ is a polynomial of
degree $k$ in $z$).~$\square $

This finding is an immediate consequence of \textit{Corollary 2.1.2}
together with the formula 
\end{subequations}
\begin{subequations}
\begin{eqnarray}
\left( c-z\right) ~F\left( z-1,-k;c;\alpha \right) +\left( 2-\alpha \right)
~z~F\left( z,-k;c;\alpha \right) &&  \notag \\
+\left( \alpha -1\right) ~z~F\left( z+1,-k;c;\alpha \right) =\left( c+\alpha
~k\right) ~F\left( z,-k;c;\alpha \right) ~, &&
\end{eqnarray}%
or equivalently (see (\ref{OpdeltaHat}))%
\begin{eqnarray}
&&\left[ \left( c-z\right) ~\hat{\delta}\left( -1\right) +\left( 2-\alpha
\right) ~z+\left( \alpha -1\right) ~z~\hat{\delta}\left( -1\right) \right]
~F\left( z,-k;c;\alpha \right)  \notag \\
&=&\left( c+\alpha ~k\right) ~F\left( z,-k;c;\alpha \right) ~,
\end{eqnarray}%
which coincides with eq. (2.8(28)) of \cite{E1953} by replacing there $z$
with $\alpha ,$ $a$ with $z$ and moreover by setting $b=-k$ so that---when $%
k $ is a nonnegative integer---$F\left( z,-k;~c;~\alpha \right) $ becomes a
polynomial of degree $k$ in $z$.

\textit{Proposition 3.3}. The $\left( N\times N\right) $-matrix 
\end{subequations}
\begin{subequations}
\label{MatWHat}
\begin{equation}
\underline{\underline{\hat{W}}}\equiv \underline{\underline{\hat{W}}}\left(
\alpha ,\beta ,\gamma ,\delta ;\underline{z}\right) =B\left( \underline{%
\underline{Z}}\right) ~\underline{\underline{\hat{\nabla}}}\left( 1;%
\underline{\zeta }\right) -B\left( -\underline{\underline{Z}}\right) ~%
\underline{\underline{\hat{\nabla}}}\left( -1;\underline{\zeta }\right) ~,
\end{equation}%
hence defined componentwise as follows in terms of the $N+4$ arbitrary
numbers $z_{n}$ and $\alpha ,~\beta ,~\gamma ,~\delta $ (see (\ref{MatZ})
and (\ref{MatNablaHat})), 
\begin{eqnarray}
&&\hat{W}_{nm}=B\left( z_{n};~\alpha ,~\beta ,~\gamma ,~\delta \right)
~\left\{ \left[ \prod\limits_{\ell =1,~\ell \neq m}^{N}\left( \frac{\left(
z_{n}+1\right) ^{2}-z_{\ell }^{2}}{z_{m}^{2}-z_{\ell }^{2}}\right) \right]
-\delta _{nm}\right\}  \notag \\
&&+B\left( -z_{n};~\alpha ,~\beta ,~\gamma ,~\delta \right) ~\left\{ \left[
\prod\limits_{\ell =1,~\ell \neq m}^{N}\left( \frac{\left( z_{n}-1\right)
^{2}-z_{\ell }^{2}}{z_{m}^{2}-z_{\ell }^{2}}\right) \right] -\delta
_{nm}\right\} ~,
\end{eqnarray}%
with 
\begin{equation}
B\left( z;~\alpha ,~\beta ,~\gamma ,~\delta \right) =\frac{\left( z+\alpha
\right) ~\left( z+\beta \right) ~\left( z+\gamma \right) ~\left( z+\delta
\right) }{2~z~\left( 2~z+1\right) }~,  \label{BWilson}
\end{equation}%
features the $N$ eigenvalues $k\left( k+\alpha +\beta +\gamma +\delta
-1\right) $ with $k$ a nonnegative integer less than $N$,%
\begin{eqnarray}
\underline{\underline{\hat{W}}}~\underline{\varphi }^{\left( k\right)
}\left( \alpha ,\beta ,\gamma ,\delta ,z_{n};\underline{z}\right)
&=&k~\left( k+\alpha +\beta +\gamma +\delta -1\right) ~\underline{\varphi }%
^{\left( k\right) }\left( \alpha ,\beta ,\gamma ,\delta ,z_{n};\underline{z}%
\right) ~,  \notag \\
k &=&0,1,...,N-1~,
\end{eqnarray}%
with the eigenvectors $\underline{\varphi }^{\left( k\right) }\left( \alpha
,\beta ,\gamma ,\delta ;\underline{z}\right) $ defined componentwise as
follows:%
\begin{equation}
\varphi _{n}^{\left( k\right) }\left( \alpha ,\beta ,\gamma ,\delta ;%
\underline{z}\right) =W_{k}\left( -z_{n}^{2};\alpha ,\beta ,\gamma ,\delta
\right) ~,  \label{EigenWilson}
\end{equation}%
where $W_{k}\left( \zeta ;\alpha ,\beta ,\gamma ,\delta \right) $ is the
Wilson polynomial of degree $k$ in $\zeta $, defined (up to an irrelevant
multiplicative constant) as follows:%
\begin{eqnarray}
&&W_{k}\left( \zeta \right) \equiv W_{k}\left( \zeta ;\alpha ,\beta ,\gamma
,\delta \right)  \notag \\
&=&\sum_{s=0}^{k}\left[ \frac{\left( -k\right) _{s}~\left( k+\alpha +\beta
+\gamma +\delta -1\right) _{s}~\left( \alpha +\mathbf{i}~z\right)
_{s}~\left( \alpha -\mathbf{i}~z\right) _{s}}{s!~\left( \alpha +\beta
\right) _{s}~\left( \alpha +\gamma \right) _{s}~\left( \alpha +\delta
\right) _{s}}\right] ~,  \notag \\
\zeta &=&-z^{2}  \label{DefWilson}
\end{eqnarray}%
(see Section 1.1 of \cite{KS1998}; hence here $\left( x\right) _{s}$ is
again the Pochhammer symbol, see (\ref{Poch}); but note that here and below
the $4$ parameters $\alpha ,\beta ,\gamma ,\delta $ are \textit{not}
required to satisfy the restrictions that are instead necessary for the
validity of some of the results reported in Section 1.1 of \cite{KS1998};
the only restrictions they must satisfy are those necessary for this
definition to make good sense). $\square $

This finding is a direct consequence of \textit{Corollary 2.1.2} (together
with \textit{Remark 2.1.3}), applied to the difference equation satisfied by
the Wilson polynomial, see eq. (1.1.6) of \cite{KS1998} (with some
appropriate changes of variables, as explained in Appendix A).

\textit{Remark 3.2}. Note again the \textit{Diophantine} character of this
result, and the fact that it implies that the $\left( N\times N\right) $%
-matrix $\underline{\underline{\hat{W}}}\left( \alpha ,\beta ,\gamma ,\delta
;\underline{z}\right) $, see (\ref{MatWHat}), which clearly depends
nontrivially on the $N+4$ numbers $z_{n}$ and $\alpha ,~\beta ,~\gamma
,~\delta $, is \textit{isospectral} for any variation of these $N+4$
parameters which leaves invariant the sum $\alpha +\beta +\gamma +\delta $. $%
\square $

A variant of this result is provided by the following

\textit{Proposition 3.4}. Let the $N$ numbers $\bar{\zeta}_{n}\equiv \bar{%
\zeta}_{n}\left( \alpha ,\beta ,\gamma ,\delta \right) =-\bar{z}_{n}^{2}$ be
the $N$ zeros of the Wilson polynomial $W_{N}\left( \zeta ;\alpha ,\beta
,\gamma ,\delta \right) $ of degree $N$ in $\zeta $ (see Section 1.1 of \cite%
{KS1998}), 
\end{subequations}
\begin{subequations}
\begin{equation}
W_{N}\left( \zeta _{n};\alpha ,\beta ,\gamma ,\delta \right)
=0,~~~n=1,...,N~,
\end{equation}%
and let the $\left( N\times N\right) $-matrix $\underline{\underline{\bar{W}}%
}\equiv \underline{\underline{\bar{W}}}\left( \alpha ,\beta ,\gamma ,\delta ;%
\underline{\bar{z}}\right) $ be defined componentwise as follows in terms of
the $N+4$ numbers $\bar{z}_{n}$ and $\alpha ,~\beta ,~\gamma ,~\delta ,$%
\begin{eqnarray}
&&\bar{W}_{nn}=-\left[ B\left( \bar{z}_{n};~\alpha ,\beta ,\gamma ,\delta
\right) +B\left( -\bar{z}_{n};~\alpha ,\beta ,\gamma ,\delta \right) \right]
\notag \\
&&+\left( \frac{4~\bar{z}_{n}}{2~\bar{z}_{n}-1}\right) ~B\left( \bar{z}%
_{n};~\alpha ,\beta ,\gamma ,\delta \right) ~\prod\limits_{\ell =1,~\ell
\neq n}^{N}\left[ \frac{\left( \bar{z}_{n}+1\right) ^{2}-\bar{z}_{\ell }^{2}%
}{\bar{z}_{n}^{2}-\bar{z}_{\ell }^{2}}\right] ~,  \notag \\
&&n=1,...,N~,
\end{eqnarray}%
\begin{eqnarray}
&&\bar{W}_{nm}=B\left( \bar{z}_{n};~\alpha ,\beta ,\gamma ,\delta \right) ~%
\left[ \prod\limits_{\ell =1,~\ell \neq m}^{N}\left( \frac{\left( \bar{z}%
_{n}+1\right) ^{2}-\bar{z}_{\ell }^{2}}{\bar{z}_{m}^{2}-\bar{z}_{\ell }^{2}}%
\right) \right]  \notag \\
&&+B\left( -\bar{z}_{n};~\alpha ,\beta ,\gamma ,\delta \right) ~\left[
\prod\limits_{\ell =1,~\ell \neq m}^{N}\left( \frac{\left( \bar{z}%
_{n}-1\right) ^{2}-\bar{z}_{\ell }^{2}}{\bar{z}_{m}^{2}-\bar{z}_{\ell }^{2}}%
\right) \right] ~,  \notag \\
n &\neq &m~,~~~n,m=1,...,N~,
\end{eqnarray}%
with $B\left( z;~\alpha ,~\beta ,~\gamma ,~\delta \right) $ defined as
above, see (\ref{BWilson}). Then this matrix features the same eigenvalues
and eigenvectors as the matrix $\underline{\underline{\hat{W}}}\left( \alpha
,\beta ,\gamma ,\delta ;\underline{z}\right) $ defined above, see \textit{%
Proposition 3.3}, except that in the definition (\ref{EigenWilson}) of the
eigenvectors the arbitrary numbers $z_{n}$ must be replaced by the $N$
numbers $\bar{z}_{n}$ such that the $N$ numbers $\bar{\zeta}_{n}=-\bar{z}%
_{n}^{2}$ are the $N$ zeros of the Wilson polynomial $W_{N}\left( \zeta
;\alpha ,\beta ,\gamma ,\delta \right) $ of degree $N$ in $\zeta $ (note
that the matrix $\underline{\underline{\bar{W}}}=\underline{\underline{\bar{W%
}}}\left( \alpha ,\beta ,\gamma ,\delta ;\underline{\bar{z}}\right) $
defined componentwise above is invariant under the exchange $\bar{z}%
_{n}\rightarrow -\bar{z}_{n},$ so it is in fact a function of the $N$
numbers $\bar{\zeta}_{n}\equiv \bar{\zeta}_{n}\left( \alpha ,\beta ,\gamma
,\delta \right) =-\bar{z}_{n}^{2}$ rather than the $N$ numbers $\bar{z}_{n}$)%
\textit{. }$\square $

For a proof of this result see Appendix A.

\textit{Remark 3.3}. This result, and analogous ones reported below (see 
\textit{Propositions 3.6} and \textit{3.9}) evoke an interesting class of 
\textit{open} problems, such as that raised by the following question.
Define the ($N\times N$)-matrix $\underline{\underline{\bar{W}}}\equiv 
\underline{\underline{\bar{W}}}\left( \alpha ,\beta ,\gamma ,\delta ;%
\underline{\bar{z}}\right) $ as in \textit{Proposition 3.4}, but assuming
that the $N$ numbers $\bar{z}_{n}$ are \textit{a priori} arbitrary; and
require that this ($N\times N$)-matrix feature the $N$ eigenvalues $k\left(
k+\alpha +\beta +\gamma +\delta -1\right) $ with $k=0,1,...,N-1.$ Does this
requirement imply that the $N$ numbers $\bar{\zeta}_{n}=-\bar{z}_{n}^{2}$
are necessarily the $N$ zeros of the Wilson polynomial $W_{N}\left( \zeta
;\alpha ,\beta ,\gamma ,\delta \right) $ of degree $N$ in $\zeta $? The
finding reported in \cite{C1982} suggests that this is \textit{not} the
case. $\square $

\textit{Proposition 3.5}. The $\left( N\times N\right) $-matrix 
\end{subequations}
\begin{subequations}
\label{MatRHat}
\begin{equation}
\underline{\underline{\hat{R}}}\equiv \underline{\underline{\hat{R}}}\left(
\alpha ,\beta ,\gamma ,\delta ;\underline{z}\right) =C\left( \underline{%
\underline{Z}}\right) ~\underline{\underline{\hat{\nabla}}}\left( 1;%
\underline{\zeta }\right) -D\left( \underline{\underline{Z}}\right) ~%
\underline{\underline{\hat{\nabla}}}\left( -1;\underline{\zeta }\right) ~,
\end{equation}%
hence defined componentwise as follows in terms of the $N+4$ arbitrary
numbers $z_{n}$ and $\alpha ,~\beta ,~\gamma ,~\delta $ (see (\ref{MatZ})
and (\ref{MatNablaHat})), 
\begin{eqnarray}
&&\hat{R}_{nm}=C\left( z_{n};~\alpha ,~\beta ,~\gamma ,~\delta \right)
~\left\{ \left[ \prod\limits_{\ell =1,~\ell \neq m}^{N}\left( \frac{\zeta
\left( z_{n}+1\right) -\zeta \left( z_{\ell }\right) }{\zeta \left(
z_{m}\right) -\zeta \left( z_{\ell }\right) }\right) \right] -\delta
_{nm}\right\}  \notag \\
&&+D\left( z_{n};~\alpha ,~\beta ,~\gamma ,~\delta \right) ~\left\{ \left[
\prod\limits_{\ell =1,~\ell \neq m}^{N}\left( \frac{\zeta \left(
z_{n}-1\right) -\zeta \left( z_{\ell }\right) }{\zeta \left( z_{m}\right)
-\zeta \left( z_{\ell }\right) }\right) \right] -\delta _{nm}\right\} ~,
\end{eqnarray}%
with%
\begin{equation}
\zeta \left( z\right) =z~\left( z+\gamma +\delta +1\right)  \label{zitaRacah}
\end{equation}%
\begin{equation}
C\left( z;~\alpha ,~\beta ,~\gamma ,~\delta \right) =\frac{\left( z+\alpha
+1\right) ~\left( z+\beta +\delta +1\right) ~\left( z+\gamma +1\right)
~\left( z+\gamma +\delta +1\right) }{\left( 2~z+\gamma +\delta +1\right)
~\left( 2~z+\gamma +\delta +2\right) }~,  \label{CRacah}
\end{equation}%
\begin{equation}
D\left( z;~\alpha ,~\beta ,~\gamma ,~\delta \right) =\frac{z~\left( z-\alpha
+\gamma +\delta \right) ~\left( z-\beta +\gamma \right) ~\left( z+\delta
\right) }{\left( 2~z+\gamma +\delta \right) ~\left( 2~z+\gamma +\delta
+1\right) }~,  \label{DRacah}
\end{equation}%
features the $N$ eigenvalues $k\left( k+\alpha +\beta +1\right) $ with $k$ a
nonnegative integer less than $N$,%
\begin{eqnarray}
\underline{\underline{\hat{R}}}~\underline{\phi }^{\left( k\right) }\left(
\alpha ,\beta ,\gamma ,\delta ,z_{n};\underline{z}\right) &=&k~\left(
k+\alpha +\beta +1\right) ~\underline{\hat{\phi}}^{\left( k\right) }\left(
\alpha ,\beta ,\gamma ,\delta ,z_{n};\underline{z}\right) ~,  \notag \\
k &=&0,1,...,N-1~,
\end{eqnarray}%
and the eigenvectors $\underline{\hat{\phi}}^{\left( k\right) }\left( \alpha
,\beta ,\gamma ,\delta ;\underline{z}\right) $ defined componentwise as
follows:%
\begin{equation}
\hat{\phi}_{n}^{\left( k\right) }\left( \alpha ,\beta ,\gamma ,\delta ;%
\underline{z}\right) =R_{k}\left( \zeta \left( z_{n}\right) ;\alpha ,\beta
,\gamma ,\delta \right) ~,  \label{EigenRacah}
\end{equation}%
where $R_{k}\left( \zeta ;\alpha ,\beta ,\gamma ,\delta \right) $ is the
Racah polynomial of degree $k$ in $\zeta $, defined (up to an irrelevant
multiplicative constant) as follows:%
\begin{eqnarray}
&&R_{k}\left( \zeta ;\alpha ,\beta ,\gamma ,\delta \right) \equiv
R_{k}\left( \zeta \left( z\right) ;\alpha ,\beta ,\gamma ,\delta \right) 
\notag \\
&=&\sum_{s=0}^{k}\left[ \frac{\left( -k\right) _{s}~\left( k+\alpha +\beta
+1\right) _{s}~\left( -z\right) _{s}~\left( z+\gamma +\delta +1\right) _{s}}{%
s!~\left( \alpha +1\right) _{s}~\left( \beta +\delta +1\right) _{s}~\left(
\gamma +1\right) _{s}}\right]  \label{DefRacah}
\end{eqnarray}%
(see Section 1.2 of \cite{KS1998}; hence here $\left( x\right) _{s}$ is
again the Pochhammer symbol, see (\ref{Poch}); but note that here and below
neither the $4$ parameters $\alpha ,\beta ,\gamma ,\delta $ nor the argument 
$\zeta $ are required to satisfy the restrictions that are instead necessary
for the validity of most of the results reported in Section 1.2 of \cite%
{KS1998}; the only restrictions they must satisfy are those necessary for
this definition to make good sense). $\square $

This finding is a direct consequence of \textit{Corollary 2.1.2} (together
with \textit{Remark 2.1.3}), applied to the difference equation satisfied by
the Racah polynomial, see eq. (1.2.5) of \cite{KS1998}; its proof is omitted
because it is analogous---\textit{mutatis mutandis}---to the proof (given in
Appendix A) of \textit{Proposition 3.3}.

\textit{Remark 3.4}. Note again the \textit{Diophantine} character of this
result, and the fact that it implies that the $\left( N\times N\right) $%
-matrix $\underline{\underline{\hat{R}}}\left( \alpha ,\beta ,\gamma ,\delta
;\underline{z}\right) $, see (\ref{MatRHat}), which clearly depends
nontrivially on the $N+4$ numbers $z_{n}$ and $\alpha ,~\beta ,~\gamma
,~\delta $, is \textit{isospectral} for any variations of these $N+4$
parameters which leaves invariant the sum $\alpha +\beta $. $\square $

\textit{Proposition 3.6}. Let the $N$ numbers $\bar{\zeta}_{n}\equiv \bar{%
\zeta}_{n}\left( \alpha ,\beta ,\gamma ,\delta \right) =\bar{z}_{n}~\left( 
\bar{z}_{n}+\gamma +\delta +1\right) $ be the $N$ zeros of the Racah
polynomial $R_{N}\left( \zeta ;\alpha ,\beta ,\gamma ,\delta \right) $ of
degree $N$ in $\zeta $ (see Section 1.2 of \cite{KS1998}), 
\end{subequations}
\begin{subequations}
\begin{equation}
R_{N}\left( \bar{\zeta}_{n};\alpha ,\beta ,\gamma ,\delta \right)
=0,~~~n=1,...,N~,
\end{equation}%
and let the $\left( N\times N\right) $-matrix $\underline{\underline{\bar{R}}%
}\equiv \underline{\underline{\bar{R}}}\left( \alpha ,\beta ,\gamma ,\delta ;%
\underline{\bar{z}}\right) $ be defined componentwise as follows in terms of
the $N+4$ numbers $\bar{z}_{n}$ and $\alpha ,~\beta ,~\gamma ,~\delta $: 
\begin{eqnarray}
\bar{R}_{nn} &=&-\left[ C\left( \bar{z}_{n}\right) +D\left( \bar{z}%
_{n}\right) \right] +2~\left( \frac{2~\bar{z}_{n}+\gamma +\delta +1}{2~\bar{z%
}_{n}+\gamma +\delta }\right) \cdot  \notag \\
&&\cdot C\left( \bar{z}_{n}\right) ~\prod\limits_{\ell =1,~\ell \neq n}^{N}%
\left[ \frac{\zeta \left( \bar{z}_{n}+1\right) -\bar{\zeta}_{\ell }}{\zeta
\left( \bar{z}_{n}-1\right) -\bar{\zeta}_{\ell }}\right] ~,  \notag \\
n &=&1,...,N~,
\end{eqnarray}%
\begin{eqnarray}
&&\bar{R}_{nm}=C\left( \bar{z}_{n};~\alpha ,~\beta ,~\gamma ,~\delta \right)
~\left[ \prod\limits_{\ell =1,~\ell \neq m}^{N}\left( \frac{\zeta \left( 
\bar{z}_{n}+1\right) -\zeta \left( \bar{z}_{\ell }\right) }{\zeta \left( 
\bar{z}_{m}\right) -\zeta \left( \bar{z}_{\ell }\right) }\right) \right] 
\notag \\
&&+D\left( \bar{z}_{n};~\alpha ,~\beta ,~\gamma ,~\delta \right) ~\left[
\prod\limits_{\ell =1,~\ell \neq m}^{N}\left( \frac{\zeta \left( \bar{z}%
_{n}-1\right) -\zeta \left( \bar{z}_{\ell }\right) }{\zeta \left( \bar{z}%
_{m}\right) -\zeta \left( \bar{z}_{\ell }\right) }\right) \right] ~, \\
n &\neq &m,~~~n,m=1,...,N~,
\end{eqnarray}%
where of course $\zeta \left( z\right) =z~\left( z+\gamma +\delta +1\right)
, $ see (\ref{zitaRacah}), and we omitted to indicate explicitly the
dependence of $C\left( \bar{z}_{n}\right) $ and $D\left( \bar{z}_{n}\right)
, $ see (\ref{CRacah}) and (\ref{DRacah}), on the $4$ parameters $\alpha
,~\beta ,~\gamma ,~\delta $. Then this matrix features the same eigenvalues
and eigenvectors as the matrix $\underline{\underline{\hat{R}}}\left( \alpha
,\beta ,\gamma ,\delta ;\underline{z}\right) $ defined above, see \textit{%
Proposition 3.1.5}, except that in the definition (\ref{EigenRacah}) of the
eigenvectors the arbitrary numbers $z_{n}$ must be replaced by the $N$
numbers $\bar{z}_{n}$ such that the $N$ numbers $\bar{\zeta}_{n}\equiv \bar{%
\zeta}_{n}\left( \alpha ,\beta ,\gamma ,\delta \right) =\bar{z}_{n}~\left( 
\bar{z}_{n}+\gamma +\delta +1\right) $ are the $N$ zeros of the Racah
polynomial $R_{N}\left( \zeta ;\alpha ,\beta ,\gamma ,\delta \right) $ of
degree $N$ in $\zeta $\textit{. }$\square $

Again, the proof of this result is omitted because it is analogous---\textit{%
mutatis mutandis}---to the proof of \textit{Proposition 3.4 }given in
Appendix A.

\textit{Remark 3.5}. Since the Wilson and the Racah polynomials are the
"highest" polynomials belonging to the Askey scheme (see for instance \cite%
{KS1998}), analogous results involving all the "lower" polynomials of the
Askey scheme can be obtained from those reported above---see \textit{%
Propositions 3.3}, \textit{3.4}, \textit{3.5} and \textit{3.6---}by
appropriate reductions. $\square $

\textit{Proposition 3.7}. The $\left( N\times N\right) $-matrix 
\end{subequations}
\begin{subequations}
\begin{equation}
\underline{\underline{\check{K}}}\left( q;\underline{z}\right) =\underline{%
\underline{Z}}~\underline{\underline{\check{\nabla}}}\left( q;\underline{z}%
\right) ~,
\end{equation}%
hence defined componentwise as follows in terms of the $N+1$ arbitrary
numbers $z_{n}$ and $q$ (see (\ref{MatZ}) and (\ref{MatNablaInvHat})), 
\begin{equation}
\check{K}_{nm}\left( q;\underline{z}\right) =\left( \frac{1}{q-1}\right)
~\left\{ \left[ \prod\limits_{\ell =1,~\ell \neq m}^{N}\left( \frac{%
q~z_{n}-z_{\ell }}{z_{m}-z_{\ell }}\right) \right] -\delta _{nm}\right\} ~,
\label{MatKInvHat}
\end{equation}%
features the $N$ eigenvalues $\left( 1-q\right) ^{k}/\left( 1-q\right) $
with $k$ the $N$ nonnegative integers less than $N$:%
\begin{equation}
\underline{\underline{\check{K}}}\left( q;\underline{z}\right) ~\underline{%
\check{v}}^{\left( k\right) }\left( q;\underline{z}\right) =\left( \frac{%
1-q^{k}}{1-q}\right) ~\underline{\check{v}}^{\left( k\right) }\left( q;%
\underline{z}\right) ~,
\end{equation}%
\begin{equation}
\check{v}_{n}^{\left( k\right) }\left( a;\underline{z}\right) =\left(
z_{n}\right) ^{k}~,~~~k=0,1,...,N-1~.~\square
\end{equation}

This result is an immediate consequence of \textit{Corollary 2.1.2},
together with the eigenvalue equation (\ref{EigenEqNablaInvHata}) and of
course the definitions (\ref{MatZ}) respectively (\ref{MatNablaInvHat}) of
the $\left( N\times N\right) $-matrix $\underline{\underline{Z}}$
respectively $\underline{\underline{\check{\nabla}}}\left( q;\underline{z}%
\right) $.

\textit{Remark 3.6}. Note again the \textit{Diophantine} character of this
result, and the fact that it implies that the $\left( N\times N\right) $%
-matrix $\underline{\underline{\hat{K}}}\left( q;\underline{z}\right) $,
which clearly depends nontrivially on the $N+1$ numbers $z_{n}$ and $q$, see
(\ref{MatKInvHat}), is \textit{isospectral} for any variations of the $N$
parameters~$z_{n}$. $\square $

\textit{Proposition 3.8}. The $\left( N\times N\right) $-matrix $\underline{%
\underline{\check{Y}}}$ defined componentwise as follows in terms of the $%
N+5 $ arbitrary numbers $z_{n}$ and $\alpha ,~\beta ,~\gamma ,~\delta ,~q$, 
\end{subequations}
\begin{subequations}
\label{MatWInvHat}
\begin{eqnarray}
&&\check{Y}_{nm}=A\left( z_{n};~\alpha ,~\beta ,~\gamma ,~\delta ;~q\right)
~\left\{ \left[ \prod\limits_{\ell =1,~\ell \neq m}^{N}\left( \frac{\zeta
\left( q~z_{n}\right) -\zeta \left( z_{\ell }\right) }{\zeta \left(
z_{m}\right) -\zeta \left( z_{\ell }\right) }\right) \right] -\delta
_{nm}\right\}  \notag \\
&&+A\left( z_{n}^{-1};~\alpha ,~\beta ,~\gamma ,~\delta ;~q\right) ~\left\{ 
\left[ \prod\limits_{\ell =1,~\ell \neq m}^{N}\left( \frac{\zeta \left(
q^{-1}~z_{n}\right) -\zeta \left( z_{\ell }\right) }{\zeta \left(
z_{m}\right) -\zeta \left( z_{\ell }\right) }\right) \right] -\delta
_{nm}\right\} ~,  \notag \\
&&
\end{eqnarray}%
with%
\begin{equation}
\zeta \equiv \zeta \left( z\right) =\frac{1}{2}~\left( z+\frac{1}{z}\right) ~
\label{zitazz}
\end{equation}%
and 
\begin{equation}
A\left( z;~\alpha ,~\beta ,~\gamma ,~\delta ;~q\right) =\frac{\left(
1-\alpha ~z\right) ~\left( 1-\beta ~z\right) ~\left( 1-\gamma ~z\right)
~\left( 1-\delta ~z\right) }{\left( 1-z^{2}\right) ~\left( 1-q~z^{2}\right) }%
~,  \label{AAskeyWilson}
\end{equation}%
features the $N$ eigenvalues $\left( q^{-k}-1\right) \left( 1-\alpha \beta
\gamma \delta q^{k-1}\right) $ with $k$ a nonnegative integer less than $N$,%
\begin{eqnarray}
\underline{\underline{\check{Y}}}~\underline{\check{\varphi}}^{\left(
k\right) }\left( \alpha ,\beta ,\gamma ,\delta ;q;\underline{z}\right)
&=&\left( q^{-k}-1\right) \left( 1-\alpha ~\beta ~\gamma ~\delta
~q^{k-1}\right) ~\underline{\check{\varphi}}^{\left( k\right) }\left( \alpha
,\beta ,\gamma ,\delta ;q;\underline{z}\right) ~,  \notag \\
k &=&0,1,...,N-1~,
\end{eqnarray}%
with the eigenvectors $\underline{\check{\varphi}}^{\left( k\right) }\left(
\alpha ,\beta ,\gamma ,\delta ;q;\underline{z}\right) $ defined
componentwise as follows:%
\begin{equation}
\check{\varphi}_{n}^{\left( k\right) }\left( \alpha ,\beta ,\gamma ,\delta
;q;z_{n}\right) =p_{k}\left( \zeta _{n};\alpha ,\beta ,\gamma ,\delta
;q;\zeta _{n}\right) ~,  \label{EigenAW}
\end{equation}%
where $p_{k}\left( \alpha ,\beta ,\gamma ,\delta ;q;\zeta \right) $ is the
Askey-Wilson polynomial of degree $k$ in $\zeta $, defined (up to an
irrelevant multiplicative constant; see eq. (3.1.1) of \cite{KS1998}) as
follows:%
\begin{eqnarray}
&&p_{k}\left( \alpha ,\beta ,\gamma ,\delta ;q;\zeta \right)  \notag \\
&=&\sum_{s=0}^{k}\left[ \frac{\left( q^{-k};q\right) _{s}~\left( \alpha
~\beta ~\gamma ~\delta ~q^{k-1};q\right) _{s}~\left[ \alpha ,\zeta ;q\right]
_{s}~q^{s}}{\left( q;q\right) _{s}~\left( \alpha ~\beta ;q\right)
_{s}~\left( \alpha ~\gamma ;q\right) _{s}~\left( \alpha ~\delta ;q\right)
_{s}}\right] ~,  \notag \\
&&\zeta =\frac{1}{2}~\left( z+\frac{1}{z}\right) ~,~~~k=0,1,2,...~,
\label{AWPol}
\end{eqnarray}%
where $\left( x;q\right) _{s}$ is the "$q$-shifted factorial", 
\begin{equation}
\left( x;q\right) _{0}=1~;~~~\left( x;q\right) _{s}=\left( 1-x\right)
~\left( 1-q~x\right) ~\left( 1-q^{2}~x\right) \cdot \cdot \cdot \left(
1-q^{s-1}~x\right) ~,~~~s=1,2,...  \label{ShiftedFac}
\end{equation}%
and%
\begin{equation}
\left[ \alpha ,\zeta ;q\right] _{s}=\left( 1-\alpha ~z;q\right) _{s}~\left(
1-\frac{\alpha }{z};q\right) _{s}=\prod\limits_{r=0}^{s-1}\left( 1-2~\alpha
~\zeta ~q^{r}~+\alpha ^{2}~q^{2r}\right)
\end{equation}%
(see Sections 0.1 and 3.1 of \cite{KS1998}; but note that here and below the 
$4$ parameters $\alpha ,\beta ,\gamma ,\delta $ are \textit{not} required to
satisfy the restrictions that are instead necessary for the validity of some
of the results reported in Section 3.1 of \cite{KS1998}; the only
restrictions they must satisfy are those necessary for this definition to
make good sense). $\square $

This finding is a direct consequence of \textit{Corollary 2.1.2} (together
with \textit{Remark 2.1.3}), applied to the difference equation satisfied by
the Askey-Wilson polynomial reading (see eq. (3.1.7) of \cite{KS1998}, with
some appropriate notational changes) 
\end{subequations}
\begin{eqnarray}
&&A\left( z\right) ~\left[ p_{k}\left( \zeta \left( q~z\right) \right)
-p_{k}\left( \zeta \right) \right] +A\left( z^{-1}\right) ~\left[
p_{k}\left( \zeta \left( \frac{z}{q}\right) \right) -p_{k}\left( \zeta
\right) \right]  \notag \\
&=&\left( q^{-k}-1\right) ~\left( 1-\alpha ~\beta ~\gamma ~\delta
~q^{k-1}\right) ~p_{k}\left( \zeta \right) ~,  \label{DiffEqAWPol}
\end{eqnarray}%
where of course $A\left( z\right) \equiv A\left( z;~\alpha ,~\beta ,~\gamma
,~\delta ;~q\right) $ and $p_{k}\left( \zeta \right) \equiv p_{k}\left(
\alpha ,\beta ,\gamma ,\delta ;q;\zeta \right) $ are defined as above, see (%
\ref{AAskeyWilson}) and (\ref{AWPol}), and---most importantly---$\zeta
\equiv \zeta \left( z\right) ,$ see (\ref{zitazz}).

\textit{Remark 3.7}. Note again the \textit{Diophantine} character of this
result, and the fact that it implies that the $\left( N\times N\right) $%
-matrix $\underline{\underline{\check{Y}}}\equiv \underline{\underline{%
\check{Y}}}\left( \alpha ,\beta ,\gamma ,\delta ;q;\underline{z}\right) $,
which clearly depends nontrivially on the $N+5$ numbers $z_{n},\alpha ,\beta
,\gamma ,\delta $ and $q$, see (\ref{MatWInvHat}), is \textit{isospectral}
for any variations of the $N+4$ parameters $z_{n},\alpha ,\beta ,\gamma
,\delta $ which keeps constant the product $\alpha \beta \gamma \delta $. $%
\square $

\textit{Proposition 3.9}. Let the $N$ numbers $\bar{z}_{n}\equiv \bar{z}%
_{n}\left( \alpha ,\beta ,\gamma ,\delta ;q;N\right) $ be such that the $N$
numbers $\bar{\zeta}_{n}\equiv \bar{\zeta}_{n}\left( \alpha ,\beta ,\gamma
,\delta ;q;N\right) =\left( \bar{z}_{n}+1/\bar{z}_{n}\right) /2$ are the $N$
zeros of the Askey-Wilson polynomial $p_{N}\left( \alpha ,\beta ,\gamma
,\delta ;q;\zeta \right) $ of degree $N$ in $\zeta $ (see (\ref{AWPol})), 
\begin{subequations}
\begin{equation}
p_{N}\left( \alpha ,~\beta ,~\gamma ,~\delta ;~q;\bar{\zeta}_{n}\right)
=0,~~~n=1,...,N~,  \label{Defznbar}
\end{equation}%
so that, up to an irrelevant multiplicative constant, 
\begin{equation}
p_{N}\left( \alpha ,~\beta ,~\gamma ,~\delta ;~q;~\zeta \right)
=\prod\limits_{k=1}^{N}\left( \zeta -\bar{\zeta}_{k}\right) ~;
\label{Pzeros}
\end{equation}%
and let the $\left( N\times N\right) $-matrix $\underline{\underline{\bar{Y}}%
}\equiv \underline{\underline{\bar{Y}}}\left( \alpha ,\beta ,\gamma ,\delta
;q;\underline{\bar{z}}\right) $ be defined componentwise as follows in terms
of the $N+5$ numbers $\bar{z}_{n}$ and $\alpha ,~\beta ,~\gamma ,~\delta
,~q, $%
\begin{eqnarray}
&&\bar{Y}_{nn}=-\left[ A\left( \bar{z}_{n}\right) +A\left( \frac{1}{\bar{z}%
_{n}}\right) \right]  \notag \\
&&\left. +\left( 1+q\right) ~\left( \frac{\bar{z}_{n}^{2}-1}{\bar{z}%
_{n}^{2}-q}\right) ~A\left( \bar{z}_{n}\right) ~\prod\limits_{\ell =1,~\ell
\neq n}^{N}\left[ \frac{\zeta \left( q~\bar{z}_{n}\right) -\zeta \left( \bar{%
z}_{\ell }\right) }{\zeta \left( \bar{z}_{m}\right) -\zeta \left( \bar{z}%
_{\ell }\right) }\right] \right\} ~,  \notag \\
n &=&1,...,N~,
\end{eqnarray}%
\begin{eqnarray}
&&\bar{Y}_{nm}=A\left( \bar{z}_{n}\right) ~\left\{ \left[ \prod\limits_{\ell
=1,~\ell \neq m}^{N}\left( \frac{\zeta \left( q~\bar{z}_{n}\right) -\zeta
\left( \bar{z}_{\ell }\right) }{\zeta \left( \bar{z}_{m}\right) -\zeta
\left( \bar{z}_{\ell }\right) }\right) \right] \right\}  \notag \\
&&+A\left( \frac{1}{\bar{z}_{n}}\right) ~\left\{ \left[ \prod\limits_{\ell
=1,~\ell \neq m}^{N}\left( \frac{\zeta \left( q^{-1}~\bar{z}_{n}\right)
-\zeta \left( \bar{z}_{\ell }\right) }{\zeta \left( \bar{z}_{m}\right)
-\zeta \left( \bar{z}_{\ell }\right) }\right) \right] \right\} ~,  \notag \\
n &\neq &m~,~~~n,m=1,...,N~,
\end{eqnarray}%
of course with $A\left( z\right) \equiv A\left( z;~\alpha ,~\beta ,~\gamma
,~\delta ;q\right) $ respectively $\zeta \left( z\right) $ defined as above,
see (\ref{AAskeyWilson}) respectively (\ref{zitazz}). Then this matrix
features the same eigenvalues and eigenvectors as the matrix $\underline{%
\underline{\check{Y}}}\equiv \underline{\underline{\check{Y}}}\left( \alpha
,\beta ,\gamma ,\delta ;q;\underline{z}\right) $ defined above, see \textit{%
Proposition 3.8}, except that in the definition (\ref{EigenAW}) of the
eigenvectors the $N$ arbitrary numbers $z_{n}$ must be replaced by the $N$
numbers $\bar{z}_{n}$\textit{. }$\square $

The proof of this result is analogous to that of \textit{Propositions 3.4}
and \textit{3.6}; it takes advantage of the relation 
\end{subequations}
\begin{eqnarray}
&&A\left( \frac{1}{\bar{z}_{n}}\right) ~\prod\limits_{\ell =1,~\ell \neq
n}^{N}\left[ \zeta \left( \frac{\bar{z}_{n}}{q}\right) -\zeta \left( \bar{z}%
_{\ell }\right) \right]  \notag \\
&=&\left( \frac{q~\bar{z}_{n}^{2}-1}{\bar{z}_{n}^{2}-q}\right) ~A\left( \bar{%
z}_{n}\right) ~\prod\limits_{\ell =1,~\ell \neq n}^{N}\left[ \zeta \left( q~%
\bar{z}_{n}\right) -\zeta \left( \bar{z}_{\ell }\right) \right]
\end{eqnarray}%
easily seen to be implied by (\ref{DiffEqAWPol}) together with (\ref%
{Defznbar}) and (\ref{Pzeros}).

\textit{Remark 3.8}. Analogous results to those reported in \textit{%
Propositions 3.8} and \textit{3.9} but involving the $q$-Racah instead of
the Askey-Wilson polynomials can be obtained in an analogous manner, or via
the relation among the polynomials of these two classes reported at the end
of Section 3.1 of \cite{KS1998}. Moreover, since the Askey-Wilson and the $q$%
-Racah polynomials are the "highest" polynomials belonging to the $q$-Askey
scheme (see for instance \cite{KS1998}), analogous results involving all the
"lower" polynomials of the $q$-Askey scheme can be obtained from those
reported above\textit{\ }by appropriate reductions. $\square $

\textit{Remark 3.9}. The results reported above involving $\left( N\times
N\right) $-matrices constructed with polynomials belonging to the Askey and $%
q$-Askey schemes have already been obtained---up to notational changes, and
some unessential restrictions on their validity---by Ryu Sasaki (private
communication, and see \cite{S2014}); for analogous results see \cite{BC2014}%
. Other recent papers reporting somewhat analogous results for the zeros of
named polynomials are listed (with no pretence to completeness) in Ref. \cite%
{Zeros}. $\square $

\textit{Remark 3.10}. Let us finally emphasize that the $\left( N\times
N\right) $-matrices $\underline{\underline{\hat{\delta}}}\left( a\right) $
respectively $\underline{\underline{\check{\delta}}}\left( q\right) $
defined componentwise by (\ref{MatdeltaHat}) respectively (\ref%
{MatdeltaInvHat}) are themselves quite \textit{remarkable}, see below the
two \textit{Remarks 4.1.1} and \textit{4.2.1} and Appendix B. And the $%
\left( N\times N\right) $-matrices $\underline{\underline{\hat{\nabla}}}%
\left( a;\underline{z}\right) $ respectively $\underline{\underline{\check{%
\nabla}}}\left( q;\underline{z}\right) $ also feature remarkable properties,
see (\ref{DeltaHatZero}) respectively (\ref{DeltaInvHatZero}). $\square $

\bigskip

\section{Proofs of the main results}

In this Section 4 we provide the missing proofs of the findings reported in
Section 2. We use of course the notation introduced above, see Sections 1
and 2. The alert and informed reader will note the analogy of these proofs
to that of (\ref{VecDerfr}) provided in Section 2.4 of \cite{C2001}; indeed
the starting point are the Lagrangian interpolational formulas (\ref%
{InterpolPol}) and (\ref{ffn}), applicable to any function $f\left( z\right) 
$ which is a polynomial in $z$ of degree less than $N$.

\bigskip

\subsection{The $\left( N\times N\right) $-matrices $\protect\underline{%
\protect\underline{Z}}$, $\protect\underline{\protect\underline{\hat{\protect%
\delta}}}\left( a;\protect\underline{z}\right) $ and $\protect\underline{%
\protect\underline{\hat{\protect\nabla}}}\left( a;\protect\underline{z}%
\right) $}

It is plain, see (\ref{InterpolPol}) and (\ref{ffn}), that 
\begin{equation}
f\left( z+a\right) =\sum_{m=1}^{N}\left[ f_{m}~p_{N-1}^{\left( m\right)
}\left( z+a\right) \right]
\end{equation}%
and that%
\begin{equation}
p_{N-1}^{\left( m\right) }\left( z+a\right) =\prod\limits_{\ell =1,~\ell
\neq m}^{N}\left( \frac{z+a-z_{\ell }}{z_{m}-z_{\ell }}\right)
~,~~~m=1,...,N~.
\end{equation}%
Hence%
\begin{equation}
f\left( z+a\right) =\sum_{m=1}^{N}\left[ f_{m}~\prod\limits_{\ell =1,~\ell
\neq m}^{N}\left( \frac{z+a-z_{\ell }}{z_{m}-z_{\ell }}\right) \right] ~,
\end{equation}%
implying, for $z=z_{n}$,%
\begin{equation}
f\left( z_{n}+a\right) =\sum_{m=1}^{N}\left[ f_{m}~\prod\limits_{\ell
=1,~\ell \neq m}^{N}\left( \frac{z_{n}+a-z_{\ell }}{z_{m}-z_{\ell }}\right) %
\right] ~.
\end{equation}%
One therefore concludes that the $N$-vector $\underline{f}\left( z+a\right)
, $ of components $f_{n}\left( z+a\right) =f\left( z_{n}+a\right) ,$ is
given by the $N$-vector formula 
\begin{subequations}
\label{far}
\begin{equation}
\underline{f}\left( z+a\right) =\underline{\underline{\hat{\delta}}}\left( a;%
\underline{z}\right) ~\underline{f}\left( z\right) ~,  \label{fa}
\end{equation}%
with the $\left( N\times N\right) $-matrix $\underline{\underline{\hat{\delta%
}}}\left( a;\underline{z}\right) $ defined by (\ref{MatdeltaHat}). Iterating 
$r$ times this formula implies 
\begin{equation}
\underline{f}\left( z+r~a\right) =\left[ \underline{\underline{\hat{\delta}}}%
\left( a;\underline{z}\right) \right] ^{r}~\underline{f}\left( z\right) ~.
\label{fra}
\end{equation}%
This coincides with (\ref{Prop21}), proving \textit{Lemma 2.1.1}.

\textit{Remark 4.1.1}. Note that the formulas (\ref{far}) clearly imply the
matrix identity 
\end{subequations}
\begin{equation}
\underline{\underline{\hat{\delta}}}\left( r~a;\underline{z}\right) =\left[ 
\underline{\underline{\hat{\delta}}}\left( a;\underline{z}\right) \right]
^{r}~,~~~r=0,1,2,...~.  \label{deltapower}
\end{equation}%
Indeed, more generally, the formulas (\ref{far}) clearly imply 
\begin{subequations}
\begin{eqnarray}
&&\underline{f}\left( z+a+b\right) =\underline{\underline{\hat{\delta}}}%
\left( a+b;\underline{z}\right) ~\underline{f}\left( \underline{z}\right) 
\notag \\
&=&\underline{\underline{\hat{\delta}}}\left( a;\underline{z}\right) ~%
\underline{\underline{\hat{\delta}}}\left( b;\underline{z}\right) ~%
\underline{f}\left( z\right) =\underline{\underline{\hat{\delta}}}\left( a;%
\underline{z}\right) ~\underline{\underline{\hat{\delta}}}\left( b;%
\underline{z}\right) ~\underline{f}\left( z\right) ~,  \label{fzab}
\end{eqnarray}%
entailing the \textit{remarkable} matrix identities 
\begin{equation}
\underline{\underline{\hat{\delta}}}\left( a;\underline{z}\right) ~%
\underline{\underline{\hat{\delta}}}\left( b;\underline{z}\right) =%
\underline{\underline{\hat{\delta}}}\left( b;\underline{z}\right) ~%
\underline{\underline{\hat{\delta}}}\left( a;\underline{z}\right) =%
\underline{\underline{\hat{\delta}}}\left( a+b;\underline{z}\right) ~.
\end{equation}%
And these formulas, together with the obvious identity $\underline{%
\underline{\hat{\delta}}}\left( 0;\underline{z}\right) =\underline{%
\underline{I}},$ with $\underline{\underline{I}}$ the $\left( N\times
N\right) $ unit matrix, also imply the \textit{remarkable} matrix identity%
\begin{equation}
\left[ \underline{\underline{\hat{\delta}}}\left( a;\underline{z}\right) %
\right] ^{-1}=\underline{\underline{\hat{\delta}}}\left( -a;\underline{z}%
\right) ~.~\square
\end{equation}

\bigskip

\subsection{The $\left( N\times N\right) $-matrices $\protect\underline{%
\protect\underline{Z}}$, $\protect\underline{\protect\underline{\check{%
\protect\delta}}}\left( q;\protect\underline{z}\right) $ and $\protect%
\underline{\protect\underline{\check{\protect\nabla}}}\left( q;\protect%
\underline{z}\right) $}

The treatment in this Section 4.2 is analogous---\textit{mutatis mutandis}%
---to that of the previous Section 4.1; hence its presentation is a bit more
terse.

The starting point are again the basic formulas of Lagrangian interpolation,
see (\ref{InterpolPol}) and (\ref{ffn}). They clearly imply 
\end{subequations}
\begin{equation}
f\left( q~z\right) =\sum_{m=1}^{N}\left[ f_{m}~p_{N-1}^{\left( m\right)
}\left( q~z\right) \right] \,,
\end{equation}%
and%
\begin{equation}
p_{N-1}^{\left( m\right) }\left( q~z\right) =\prod\limits_{\ell =1,~\ell
\neq m}^{N}\left( \frac{q~z-z_{\ell }}{z_{m}-z_{\ell }}\right)
~,~~~m=1,...,N~;
\end{equation}%
hence%
\begin{equation}
f\left( q~z\right) =\sum_{m=1}^{N}\left[ f_{m}~\prod\limits_{\ell =1,~\ell
\neq m}^{N}\left( \frac{q~z-z_{\ell }}{z_{m}-z_{\ell }}\right) \right] ~,
\end{equation}%
implying, for $z=z_{n}$,%
\begin{equation}
f\left( q~z_{n}\right) =\sum_{m=1}^{N}\left[ f_{m}~\prod\limits_{\ell
=1,~\ell \neq m}^{N}\left( \frac{q~z_{n}-z_{\ell }}{z_{m}-z_{\ell }}\right) %
\right] ~.
\end{equation}%
One therefore concludes that the $N$-vector $\underline{f}\left( q~z\right)
, $ of components $f_{n}\left( q~z\right) =f\left( q~z_{n}\right) ,$ is
given by the $N$-vector formula 
\begin{subequations}
\label{fqr}
\begin{equation}
\underline{f}\left( q~z\right) =\underline{\underline{\check{\delta}}}\left(
q;\underline{z}\right) ~\underline{f}\left( z\right) ~,
\end{equation}%
with the $\left( N\times N\right) $-matrix $\underline{\underline{\check{%
\delta}}}\left( a;\underline{z}\right) $ defined by (\ref{MatdeltaInvHat}).
Iterating $r$ times this formula one gets 
\begin{equation}
\underline{f}\left( q^{r}~z\right) =\left[ \underline{\underline{\check{%
\delta}}}\left( q;\underline{z}\right) \right] ^{r}~\underline{f}\left(
z\right) ~,~~~r=1,2,...~.
\end{equation}%
This coincides with (\ref{Prop21}), proving \textit{Lemma 2.2.1}.

\textit{Remark 4.2.1}. Note that the formulas (\ref{fqr}) clearly imply the
matrix identity%
\begin{equation}
\underline{\underline{\check{\delta}}}\left( r^{q};\underline{z}\right) =%
\left[ \underline{\underline{\check{\delta}}}\left( q;\underline{z}\right) %
\right] ^{r}~,~~~r=0,1,2,...~.
\end{equation}%
Indeed, more generally, the formulas (\ref{fqr}) clearly imply 
\end{subequations}
\begin{subequations}
\begin{eqnarray}
&&\underline{f}\left( pqz\right) =\underline{\underline{\check{\delta}}}%
\left( pq;\underline{z}\right) ~\underline{f}\left( \underline{z}\right) 
\notag \\
&=&\underline{\underline{\check{\delta}}}\left( p;\underline{z}\right) ~%
\underline{\underline{\check{\delta}}}\left( q;\underline{z}\right) ~%
\underline{f}\left( z\right) =\underline{\underline{\check{\delta}}}\left( q;%
\underline{z}\right) ~\underline{\underline{\check{\delta}}}\left( p;%
\underline{z}\right) ~\underline{f}\left( z\right) ~,
\end{eqnarray}%
entailing the \textit{remarkable} matrix identities 
\begin{equation}
\underline{\underline{\check{\delta}}}\left( p;\underline{z}\right) ~%
\underline{\underline{\check{\delta}}}\left( q;\underline{z}\right) =%
\underline{\underline{\check{\delta}}}\left( q;\underline{z}\right) ~%
\underline{\underline{\check{\delta}}}\left( p;\underline{z}\right) =%
\underline{\underline{\check{\delta}}}\left( pq;\underline{z}\right) ~.
\end{equation}%
And these formulas, together with the obvious identity $\underline{%
\underline{\check{\delta}}}\left( 1;\underline{z}\right) =\underline{%
\underline{I}},$ with $\underline{\underline{I}}$ the $\left( N\times
N\right) $ unit matrix, also imply the remarkable identity%
\begin{equation}
\left[ \underline{\underline{\check{\delta}}}\left( q;\underline{z}\right) %
\right] ^{-1}=\underline{\underline{\check{\delta}}}\left( q^{-1};\underline{%
z}\right) ~.~\square
\end{equation}

\bigskip

\section{Outlook}

In this last section we mention tersely possible future developments.

A possibility is to investigate finite-dimensional representations of
difference operators which are \textit{exact} in other functional spaces:
for instance in functional spaces spanned by other seed functions than
polynomials (of degree less than $N$), possibly also including functions of
more than a single dependent variable. For previous developments in these
directions see for instance \cite{C1997}, \cite{C2001}.

Another possibility is to identify many more \textit{remarkable} matrices
than the representative examples reported in Section 3; such as those
discussed in the various subsections of Section 2.4.5 (entitled "Remarkable
matrices and identities") and in Appendix D (entitled "Remarkable matrices
and related identities") of \cite{C2001}, and in other papers referred to
there.

Yet another possibility is to identify \textit{solvable} nonlinear dynamical
systems, for instance by extending to exact finite-dimensional
representations of \textit{difference} operators the techniques based on
exact finite-dimensional representations of \textit{differential} operators,
as described in Section 2.5 (entitled "Many-body problems on the line
solvable via techniques of exact Lagrangian interpolation") of \cite{C2001}.

\bigskip

\section{Acknowledgements}

It is a pleasure to thank the organizers of the CRM-ICMAT Workshop on
"Exceptional orthogonal polynomials and exact solutions in mathematical
physics" (Segovia, Spain, 7--12 July 2014), and in addition professor Ryu
Sasaki, because the derivation and presentation of the findings reported in
this paper were motivated by discussions with him initiated at that meeting
and continued via a few e-mail exchanges.

\bigskip

\section{Appendix A}

In this Appendix A we tersely outline the derivation of \textit{Propositions
3.3} and \textit{3.4}.

The starting point to prove \textit{Propositions 3.3} is the difference
equation satisfied by the Wilson polynomial $W_{k}\left( \zeta ;\alpha
,\beta ,\gamma ,\delta \right) $ of degree $k$ in $\zeta $ (see eq. (1.1.6)
of \cite{KS1998}; and note that the validity of this relation for Wilson
polynomials does not require any limitation on the 4 parameters $\alpha
,\beta ,\gamma ,\delta $ other than those required in order that their
definition (\ref{DefWilson}) make good sense). We conveniently write this
difference equation as follows, via some trivial notational changes
(including the relations $x=\mathbf{i}z$ and $\zeta =x^{2}=-z^{2}$, and the
omission of the explicit indication of dependence on the $4$ parameters $%
\alpha ,~\beta ,~\gamma ,~\delta $): 
\end{subequations}
\begin{eqnarray}
&&\left[ B\left( z\right) ~\hat{\nabla}\left( 1\right) -B\left( -z\right) ~%
\hat{\nabla}\left( -1\right) \right] ~W_{k}\left( -z^{2}\right)  \notag \\
&=&k~\left( k+\alpha +\beta +\gamma +\delta -1\right) ~W_{k}\left(
-z^{2}\right) ~,~~~k=0,1,2,...~,  \label{BDeltaHat}
\end{eqnarray}%
with $B\left( z\right) $ defined as above, see (\ref{BWilson}), and the
operators $\hat{\nabla}\left( \pm 1\right) $ acting on functions of the
variable $z$ as follows (see (\ref{OpNablaHat})): 
\begin{equation}
\hat{\nabla}\left( \pm 1\right) ~f\left( z\right) =\pm \left[ f\left( z\pm
1\right) -f\left( z\right) \right] ~.
\end{equation}%
It is plain that the operator 
\begin{equation}
\hat{D}=B\left( z\right) ~\hat{\nabla}\left( 1\right) -B\left( -z\right) ~%
\hat{\nabla}\left( -1\right) ~,
\end{equation}%
when acting on functions of the variable $\zeta =-z^{2},$ yields functions
of this variable $\zeta $ (not of $z$), hence we can consider the equation (%
\ref{BDeltaHat}) as an eigenvalue equation,%
\begin{equation}
\hat{D}~W_{k}\left( \zeta \right) =k~\left( k+\alpha +\beta +\gamma +\delta
-1\right) ~W_{k}\left( \zeta \right) ~,
\end{equation}%
satisfied by polynomials $W_{k}\left( \zeta \right) $ of degree $k$ in the
variable $\zeta $.

The validity of \textit{Proposition 3.3 }is then an immediate consequence of 
\textit{Corollary 2.1.2}, together with \textit{Remark 2.1.3}.

Moreover, let the $N$ zeros of the polynomial $W_{N}\left( -z^{2}\right)
\equiv W_{N}\left( \zeta \right) $ be denoted as $\bar{\zeta}_{j}=-\bar{z}%
_{j}^{2},$ $j=1,...,N$, so that (up to an irrelevant multiplicative constant)%
\begin{equation}
W_{N}\left( -z^{2}\right) =\prod\limits_{j=1}^{N}\left( -z^{2}+\bar{z}%
_{j}^{2}\right) =\prod\limits_{j=1}^{N}\left( \zeta -\bar{\zeta}_{j}\right)
~.
\end{equation}%
Then (\ref{BDeltaHat}) with $k=N$ and $z=\bar{z}_{n}$, $n=1,...,N$, implies 
\begin{subequations}
\begin{eqnarray}
&&B\left( \bar{z}_{n}\right) ~\prod\limits_{j=1}^{N}\left[ \left( \bar{z}%
_{n}+1\right) ^{2}-\bar{z}_{j}^{2}\right] +B\left( -\bar{z}_{n}\right)
~\prod\limits_{j=1}^{N}\left[ \left( \bar{z}_{n}-1\right) ^{2}-\bar{z}%
_{j}^{2}\right] =0~,  \notag \\
&&n=1,...,N~,
\end{eqnarray}%
entailing%
\begin{eqnarray}
\left( 2~\bar{z}_{n}+1\right) ~B\left( \bar{z}_{n}\right)
~\prod\limits_{\ell =1,~\ell \neq n}^{N}\left[ \left( \bar{z}_{n}+1\right)
^{2}-\bar{z}_{\ell }^{2}\right] &&  \notag \\
+\left( -2~\bar{z}_{n}+1\right) B\left( -\bar{z}_{n}\right)
~\prod\limits_{\ell =1,~\ell \neq n}^{N}\left[ \left( \bar{z}_{n}-1\right)
^{2}-\bar{z}_{\ell }^{2}\right] &=&0
\end{eqnarray}%
hence%
\begin{eqnarray}
B\left( -\bar{z}_{n}\right) ~\prod\limits_{\ell =1,~\ell \neq n}^{N}\left[
\left( \bar{z}_{n}-1\right) ^{2}-\bar{z}_{\ell }^{2}\right] &=&\left( \frac{%
2~\bar{z}_{n}+1}{2~\bar{z}_{n}-1}\right) \cdot  \notag \\
\cdot B\left( \bar{z}_{n}\right) ~\prod\limits_{\ell =1,~\ell \neq n}^{N}%
\left[ \left( \bar{z}_{n}+1\right) ^{2}-\bar{z}_{\ell }^{2}\right] ~,~~~n
&=&1,...,N~.  \notag
\end{eqnarray}

Now one observes that \textit{Proposition 3.3} holds for \textit{any}
arbitrary assignment of the $N$ numbers $z_{n}$; hence it holds in
particular for the assignment $z_{n}=\bar{z}_{n}$. And with this
assignment---via the last formula written above---\textit{Proposition 3.3}
clearly becomes \textit{Proposition 3.4}, which is thereby proven.

\bigskip

\section{Appendix B}

In this Appendix B we list two \textit{remarkable} identities satisfied by
the two $\left( N\times N\right) $-matrices $\underline{\underline{\hat{%
\delta}}}\left( a;\underline{z}\right) $ and $\underline{\underline{\check{%
\delta}}}\left( q;\underline{z}\right) $. Their proof is an immediate
consequence of the identities satisfied by the two corresponding operators $%
\hat{\delta}\left( a\right) $ and $\check{\delta}\left( q\right) ,$ the
validity of which is quite obvious. 
\end{subequations}
\begin{equation}
\underline{\underline{\check{\delta}}}\left( q^{-1};\underline{z}\right) ~%
\underline{\underline{\hat{\delta}}}\left( a;\underline{z}\right) ~%
\underline{\underline{\check{\delta}}}\left( q;\underline{z}\right) =%
\underline{\underline{\hat{\delta}}}\left( q~a;\underline{z}\right) ~;
\end{equation}%
\begin{equation}
\underline{\underline{\check{\delta}}}\left( q^{-1};\underline{z}\right) ~%
\underline{\underline{\hat{\delta}}}\left( a;\underline{z}\right) ~%
\underline{\underline{\check{\delta}}}\left( q;\underline{z}\right) ~%
\underline{\underline{\hat{\delta}}}\left( b;\underline{z}\right) =%
\underline{\underline{\hat{\delta}}}\left( a+q^{-1}~b;\underline{z}\right) ~.
\end{equation}%
Other analogous identities can be obviously obtained.

It is moreover plain that there holds the matrix-vector eigenvalue equation 
\begin{subequations}
\begin{equation}
\underline{\underline{\check{\delta}}}\left( q;\underline{z}\right) ~\left( 
\underline{\underline{Z}}~\underline{u}\right) =q^{k}~\left( \underline{%
\underline{Z}}~\underline{u}\right) ~,~~~k=0,1,...,N-1~,
\end{equation}%
implying%
\begin{equation}
\det \left[ \underline{\underline{\check{\delta}}}\left( q;\underline{z}%
\right) \right] =q^{N\left( N-1\right) /2}~,~~~\text{trace}\left[ \underline{%
\underline{\check{\delta}}}\left( q;\underline{z}\right) \right] =\frac{%
1-q^{N}}{1-q}~.
\end{equation}

Here of course the two $\left( N\times N\right) $-matrices $\underline{%
\underline{\hat{\delta}}}\left( a;\underline{z}\right) $ and $\underline{%
\underline{\check{\delta}}}\left( q;\underline{z}\right) $ are defined
componentwise by (\ref{Matdeltas}) in terms of the $N$ components $z_{n}$ of
the $N$-vector $\underline{z}$ (which are $N$ \textit{arbitrary} numbers,
except for the restriction to be all different among themselves), and of the
arbitrary parameters $a$ and $q;$ while $\underline{\underline{Z}}=$diag$%
\left[ z_{n}\right] $ and the $N$-vector $\underline{u}$ has all components
equal to unity, $\underline{u}=\left( 1,1,...,N\right) $.

\bigskip\

\end{subequations}

\end{document}